\documentclass[12pt]{article}
\usepackage{amsmath,amssymb,graphicx}
\usepackage{floatflt}
\usepackage [small] {caption}

\usepackage{amsfonts}
\usepackage{color}

\usepackage[T2A]{fontenc}
\usepackage[cp1251]{inputenc}

\numberwithin{equation}{section}
\newcommand{\ra}{\rightarrow}
\newcommand{\bq}{\begin{eqnarray}}
\newcommand{\eq}{\end{eqnarray}}
\newcommand{\ov}{\overline}

\renewcommand{\mathbf}{\vec}
\newcommand{\beq}{\begin{equation}}
\newcommand{\eeq}{\end{equation}}

\begin{document}
\vspace*{4cm}
\begin{center}{\bf  Hard two photon processes \boldmath {$\gamma\gamma\to {\ov M}_2 M_1$} in QCD}
\end{center}
\vspace*{1cm}
\begin{center}
\bf{Victor L. Chernyak}\end{center}
\begin{center}(e-mail: v.l.chernyak@inp.nsk.su) \end{center}
\vspace*{1mm}
\begin{center} Budker Institut of Nuclear Physics SB RAS and Novosibirsk State University,\\ 630090 Novosibirsk, Russia
\end{center}
\vspace*{1cm}

\begin{center}
Talk given at the Taipei International Workshop \\ "QCD in two photon processes"\\ 2 - 4 \,  October 2012,\,\, Taipei,\,\, Taiwan
\end{center}
\vspace*{1cm}

\begin{center}\bf{Abstract}\end{center}

A short review of leading term QCD predictions vs those of the handbag model for large angle cross sections $\gamma\gamma\to {\ov P}_2 P_1$ ($P$ is the pseudoscalar meson $\pi^{\pm,\,o},\,K^{\pm,\,o},\,
\eta$), and for $\gamma\gamma\to {\ov V}_2 V_1$ ($V$ is the neutral vector meson $\rho^o,\,\omega,\,\phi$), in comparison with Belle Collaboration measuments

\newpage
\begin{center}{\bf Leading term QCD predictions}\end{center}

The general approach to calculations of hard exclusive processes in QCD was developed in \cite{cz1} and \cite{ER} (the operator expansions and resummation of Feynman diagrams in the covariant perturbation theory), and in \cite{LB} (the resummation of Feynman diagrams in the non-covariant light front perturbation theory in the special axial gauge and in the basis of free on mass-shell quarks and gluons).
The review is \cite{cz-r}.\\

In particular, the calculation of the large angle scattering amplitudes $\gamma\gamma\to M_2 M_1$ was considered in \cite{BL}  (for symmetric meson wave functions only, $\phi_M(x)=\phi_M(1-x)$) and \cite{Maurice} (for general wave functions), see Fig.1.\\
\vspace*{3mm}

\begin{minipage}{0.65\textwidth}
\hspace*{-1cm}
\includegraphics[width=0.85\textwidth]{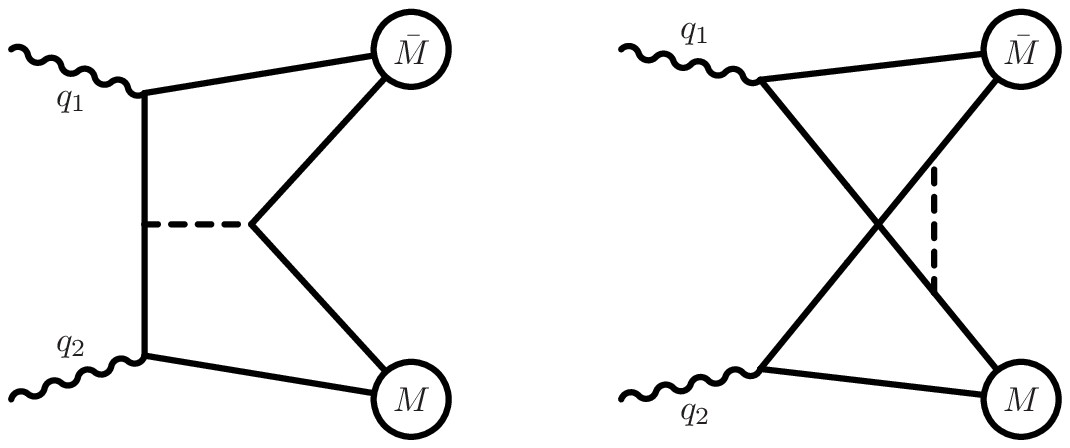}
\end{minipage}
\begin{minipage}{0.3\textwidth}
\hspace*{3mm} Fig.1 \,\, Two typical lowest order Feynman diagrams for the leading term hard QCD contributions to the large angle amplitude $\gamma\gamma\ra {\ov M}_2 M_1$ (the dashed line is the hard
gluon exchange).
\end{minipage}
\vspace*{6mm}

The leading term QCD expressions for cross sections look as (the example is given for $\gamma\gamma\to K^+K^-$)\,:
\beq
\frac{d\sigma(\gamma\gamma\ra M^{\dagger} M)}{d\cos\theta}=\frac{1}{32\pi W^2}\,\frac{1}{4}\sum_{
\lambda_1, \lambda_2=\pm 1}\,\,\Bigl | M_{\lambda_1\lambda_2}\Bigr |^2\,, \nonumber
\eeq
\beq
M^{(lead)}_{\lambda_1\lambda_2}(W,\theta)=\frac{64\pi^2}{9W^2}\,\alpha \,{\ov \alpha}_s \,f_P^2\int_0^1 dx_s\, \phi_{P}(x_s)\int_0^1 dy_s\, \phi_{P}(y_s)\,T_{\lambda_1\lambda_2}
(x_s,\,y_s,\,\theta)\,,\nonumber
\eeq
\beq
T_{++}=T_{--}=(e_u-e_s)^2\,\frac{1}{\sin^2\theta}\,\frac{A}{D}\,,\quad T_{+-}=T_{-+}\,, \nonumber
\eeq
\beq
T_{+-}=\frac{1}{D}\Biggl [ \frac{(e_u-e_s)^2}{\sin^2\theta}(1-A)+e_u e_s \frac
{A C}{A^2-B^2\cos^2\theta}+\frac{(e_u^2-e_s^2)}{2}(x_u-y_s)\Biggr ],\hspace*{2cm} (1)\nonumber
\eeq
\beq
A=(x_sy_u+x_uy_s),\,\, B=(x_sy_u-x_uy_s),\,\, C=(x_sx_u+y_sy_u),\,\, D=x_u x_sy_u y_s\,,\nonumber
\eeq
where: $x$ is the meson momentum fraction carried by quark inside the meson,
$x_s+x_u=1\,,\quad e_u=2/3, \quad e_s=e_d=-1/3\,,\,\,   \phi_P(x)$ is the leading twist pseudoscalar meson wave function (= distribution amplitude), $f_P$ are the decay constants\,:\,
$f_{\pi}\simeq 131\,{\rm MeV}\,,\,\, f_K\simeq 161\,{\rm MeV}.$ \\

The leading contribution to  $d\sigma(\gamma\gamma\to\pi^+\pi^-)$ can be written as\,:
\beq
\frac{s^3}{16\pi\alpha^2}\,\frac{d\sigma(\gamma\gamma\ra\pi^+\pi^-)}{d |\cos\theta |}
\equiv \frac{|\Phi^{(eff)}_{\pi}(s,\theta)|^2}{\sin ^4 \theta}
=\frac{|s F_{\pi}^{(lead)}(s)|^2}{\sin ^4 \theta}|1- \upsilon(\theta)|^2,\hspace*{3cm}(2)\nonumber
\eeq
where
$F_{\pi}^{(lead)}(s)$ is the leading term of the pion form factor \,:
\beq
|s F^{(lead)}_{\pi}(s)|= \frac{8\pi\,{\ov\alpha}_s}{9}\,\Bigl
|f_{\pi}\int_0^1 \frac{dx}{x}\,\phi_{\pi}(x,\,{\ov \mu})\Bigr |^2\,,\hspace*{7.5cm}(3)\nonumber
\eeq
and $\upsilon(\theta)$ is due to the term $\sim AC$ in (1). We will compare below the
predictions of two frequently used models for $\phi_{\pi}(x): \phi^{asy}(x)=6x(1-x)$ and $\phi_{\pi}^{CZ}(x,\mu_o)=30x(1-x)(2x-1)^2,\, \mu_o\sim 1{\rm GeV}$\,~\cite{cz2}.

While the numerical value of $|s F^{(lead)}_{\pi}(s)|$ is highly sensitive to the form of $\phi_{\pi}(x,\ov\mu)$, the function $\upsilon(\theta)$ is only weakly dependent of $\theta$ at $|\cos\theta|<0.6$ and, as emphasized in \cite{BL}, is weakly sensitive to the form of $\phi_{\pi}
(x,\ov\mu)$. For the above two very different pion wave functions,\,\, ${\ov\upsilon}(\theta)\simeq 0.12$.
\vspace*{6mm}

\begin{minipage}{0.5\textwidth}\hspace*{-5mm}
\includegraphics[width=0.85\textwidth]{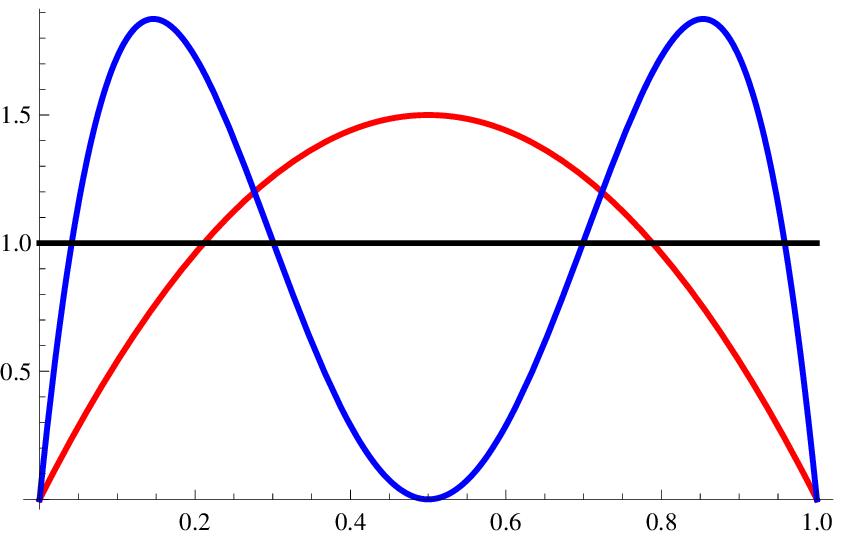}
\end{minipage}
\begin{minipage}{0.35\textwidth}
\vspace{0.5mm}
Red line\, : the asymptotic wave function\,\, $\phi_{\pi}^{\rm asy}(x)=6x(1-x)$\\
\vspace{-4mm}

Blue line\, :  the CZ wave function (at the low scale normalization point
$\mu_o\sim 1\,GeV$)

$\phi^{\rm CZ}_{\pi}(x,\mu_o)=30x(1-x)(2x-1)^2$\\
\vspace{-3mm}

Black line\, :  the flat wave function\,\, $\phi_{\pi}(x,\mu_o)=1$\\

\vspace*{-2mm}

\hspace*{1cm} $\int_0^1 dx\, \phi_{\pi}(x, \mu/\Lambda_{QCD})=1$
\end{minipage}

Fig.2\quad Different models for the leading twist \\ \hspace*{3cm} pion wave function

\vspace*{2mm}

\begin{minipage}{0.5\textwidth}\hspace*{-5mm}
\includegraphics[width=0.87\textwidth]{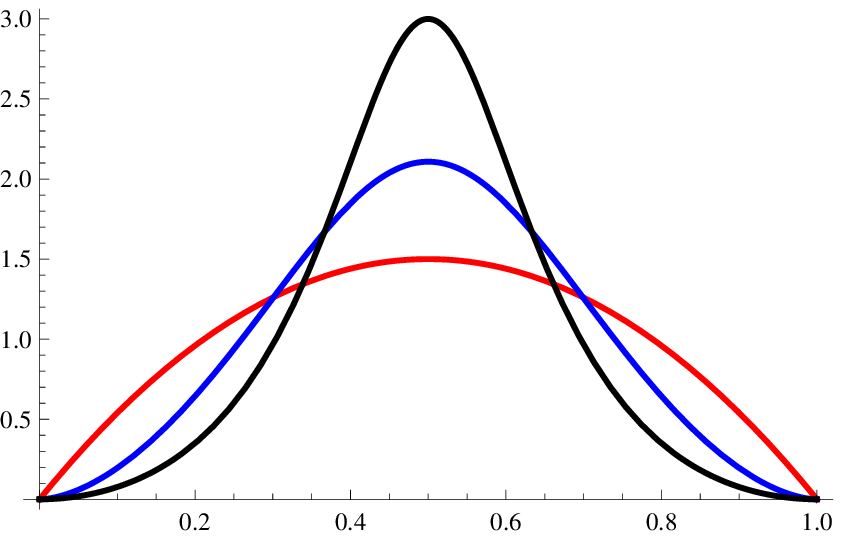}
\end{minipage}
\begin{minipage}{0.45\textwidth}
\vspace*{13mm}

Fig.3\quad The shape of the quarkonium model wave function $\phi(x,{\rm v}^2)$. Red line:\, ${\rm v}^2=1$ - massless quarks (asymptotic). Blue line:\, ${\rm v}^2=0.3$ - charmonium. Black line:\, ${\rm v}^2=0.1$ - bottomonium \\
\vspace*{2mm}

\hspace*{2cm} $\int_0^1 dx \phi_{\pi}(x, {\rm v}^2)=1$.\\

\vspace*{1mm}

\hspace*{20mm} {\it \bf  The heavier is quark} \\ \hspace*{10mm} {\it \bf the narrower is wave function}
\end{minipage}

\vspace*{10mm}

\begin{minipage}[c]{.5\textwidth}\hspace*{-5mm}
\includegraphics[trim=0mm 0mm 0mm 0mm, width=0.9\textwidth,clip=true]{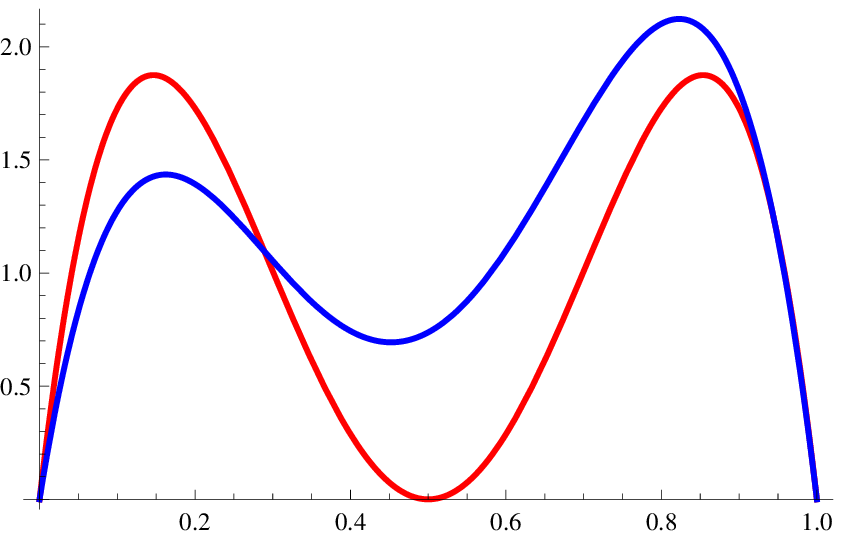}
\end{minipage}
\begin{minipage}[c]{.55\textwidth}
\vspace*{10mm}

Red line\,: \, the model pion wave function\\
\vspace*{0.5mm}

$\hspace*{1.5cm}\phi_{\pi}(x)=30\,x_d x_u(x_d-x_u)^2$\\
\vspace*{5mm}

Blue line\,:\, the model kaon wave function\\
\vspace*{0.5mm}

$\hspace*{-7mm}\phi_K(x)=30\,x_sx_u\Bigl [0.6(x_s-x_u)^2+0.08+0.1(x_s-x_u)\Bigr ]$\\
\vspace*{0.1mm}
\end{minipage}
\vspace*{2mm}

\hspace*{1mm} Fig.4\quad The kaon wave function is somewhat narrower than the pion one and asymmetric : \\
\, the s-quark in the K-meson carries a larger part of the momentum fraction than the u-quark
\newpage
\vspace*{1mm}

Therefore, the leading term QCD predictions for charged mesons $\pi^+\pi^-$ and $K^+K^-$ look as $d\sigma/d\cos\theta\sim {\ov\alpha}^{\,2}_s/(W^6\sin^4\theta)$. The recent data from Belle \cite{Nakaz} agree with the $\sim 1/\sin^4\theta $ dependence at $W\geq 3\,GeV$, while the angular distribution is somewhat steeper at lower energies. The energy dependence at $2.4\,{\rm GeV} < W < 4.1\, {\rm GeV}$ was fitted in \cite{Nakaz} as: $\sigma_o(\pi^+\pi^-, |\cos\theta|<0.6)=\int_{-0.6}^{+0.6}dz (d\sigma/dz) \sim W^{-n}\,,\,\, \,n=(7.9\pm 0.4\pm 1.5)$ for $\pi^+\pi^-$, and $n=(7.3\pm 0.3\pm 1.5)$ for $K^+K^-$. However, the overall value $n\simeq 6 $ is also acceptable, see Fig.5\,.

As for the absolute normalization, the $\pi^+\pi^-$ data are fitted \cite{Nakaz} with\,:\\
$\hspace*{4cm}|\Phi_{\pi}^{(eff)} (s,\theta)|=(0.503\pm 0.007\pm 0.035)\,{\rm GeV}^2$.
\vspace*{0.5mm}

Clearly, in addition to the leading terms $A^{(lead)}$, this experimental value includes also all power corrections $\delta A$ to the $\gamma\gamma\ra \pi^+\pi^- $ amplitudes $A=A^{(lead)}+\delta A$. These are different from corrections $\delta F_{\pi}$ to the genuine pion form factor $F_{\pi}=F_{\pi}^{(lead)}
+\delta F_{\pi}$. So, the direct connection between the leading terms of $d\sigma(\pi^+\pi^-)$
and $|F_{\pi}|^2$ in (2),(3) does not hold on account of corrections.
\vspace*{15mm}

\begin{minipage}[c]{.5\textwidth}
\includegraphics[trim=0mm 0mm 0mm 0mm, width=0.75\textwidth,clip=true]{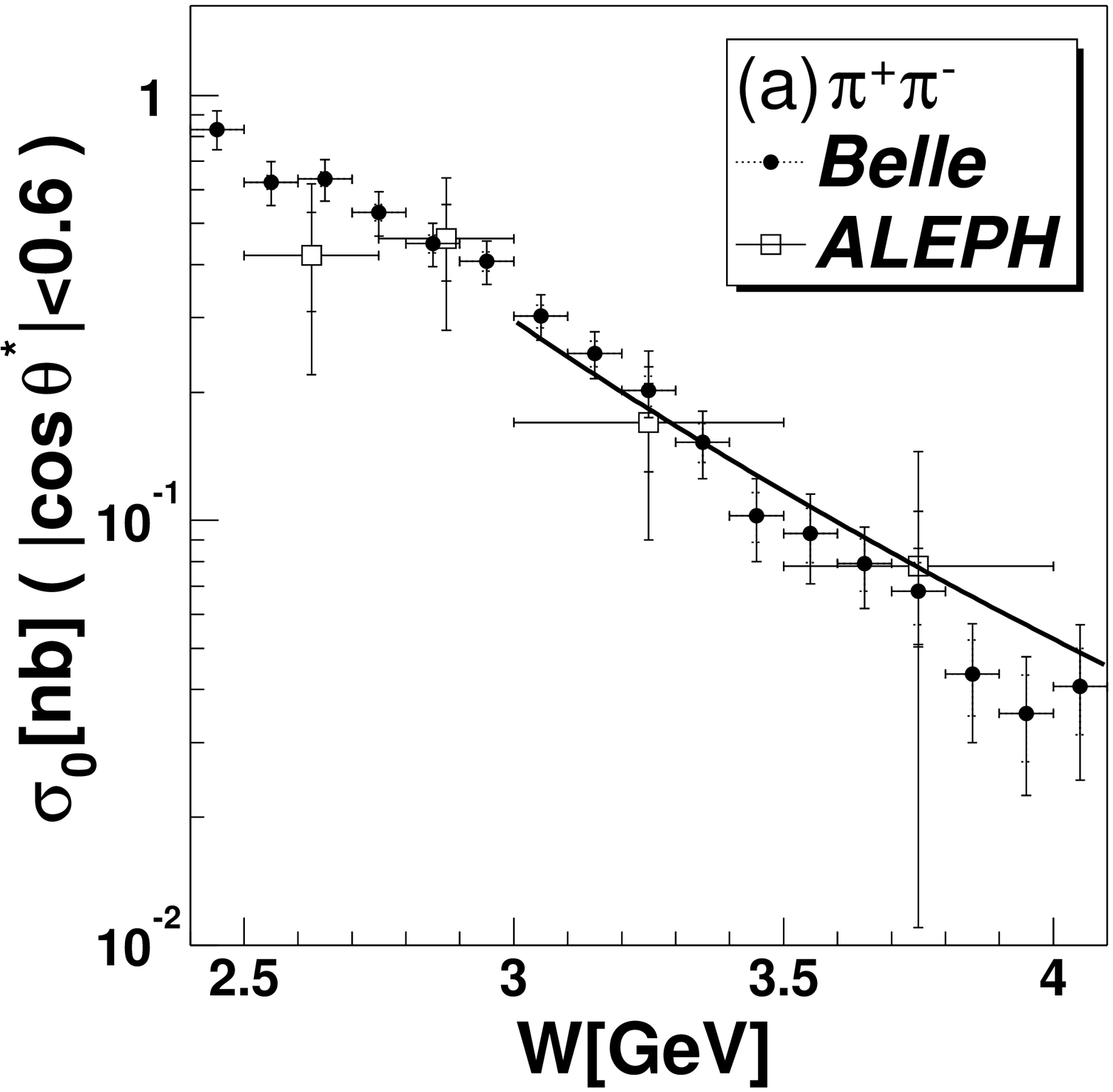}
\end{minipage}~
\begin{minipage}[c]{.5\textwidth}
\includegraphics[trim=0mm 0mm 0mm 0mm, width=0.75\textwidth,clip=true]{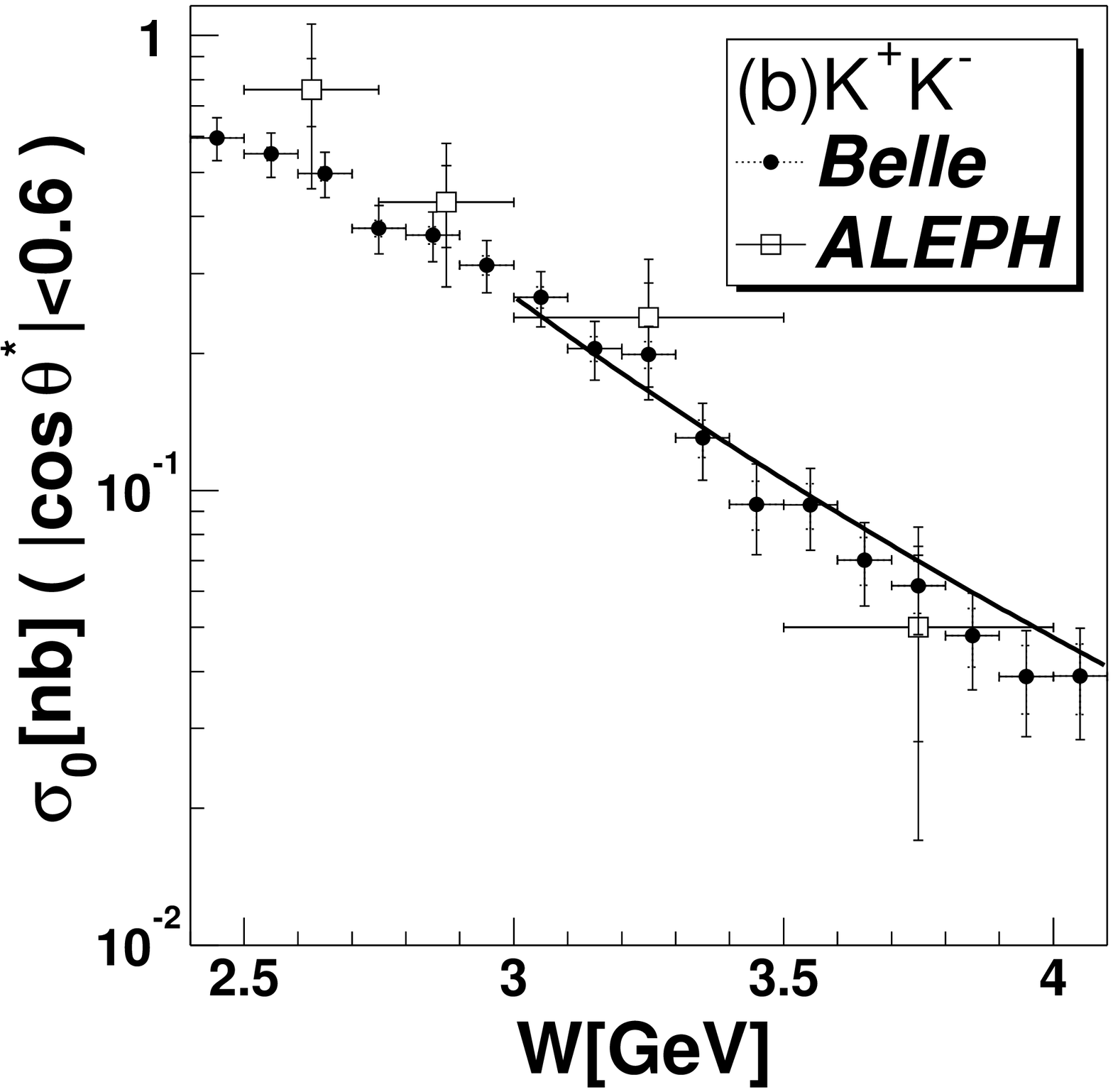}
\end{minipage}

\begin{minipage}[c]{0.9\textwidth}
\vspace{2mm}

\hspace*{4cm}
\includegraphics[width=0.55\textwidth]{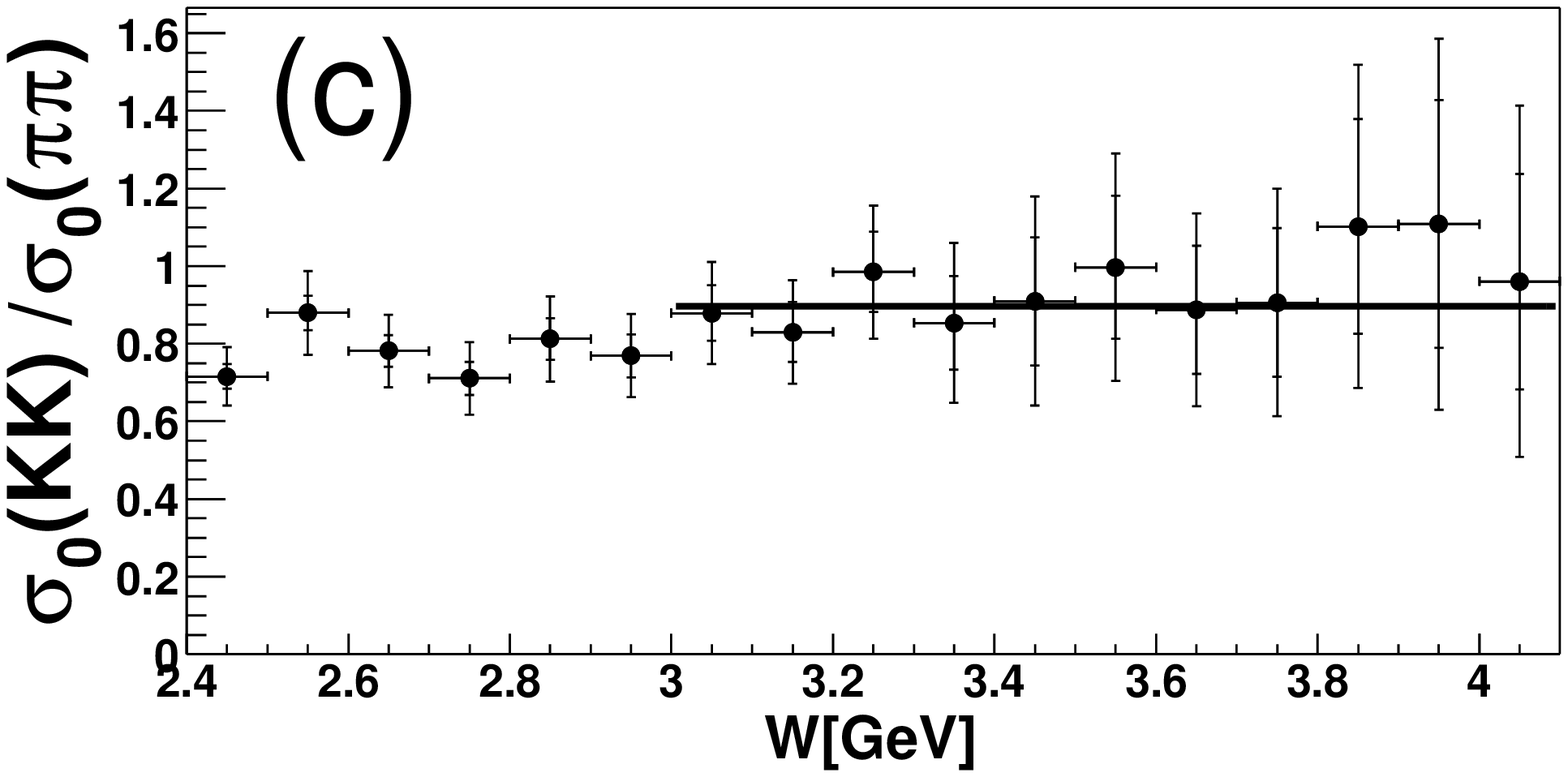}
\end{minipage}

\vspace*{1mm}
\hspace*{3mm} Fig.5\,\, a,\,b)\, Cross\, sections\,\, $\sigma_{o}(\gamma\gamma\to \pi^+\pi^-)$ and
$\sigma_{o}(\gamma\gamma\to K^+ K^-)$ integrated over \\ the angular region $|\cos\theta|<0.6$, together
with the  $\sim (1/W)^6$ dependence line.\\
\vspace*{3mm}
\hspace*{8mm} Fig.5\,  c)\, the cross section ratio\, $R_{\rm exp}=\sigma_{o}(K^+K^-)/\sigma_{o}(\pi^+
\pi^-)\simeq 0.9$.\\ Compare $R_{\rm exp}\simeq 0.9$ with the naive prediction for $\phi_{\pi}(x)=\phi_K (x)\,: \,R=(f_K/f_{\pi})^4\simeq 2.3$\,.\\

The experimental value $|\Phi_{\pi}^{(eff)}(s,\theta)|=|1- \upsilon(\theta)||s F_{\pi}^{(lead)}(s)|\simeq 0.88 | F_{\pi}^{(lead)}(s)|\simeq 0.5\,{\rm GeV}^2$ can be compared with $0.88\cdot |s\, F_{\pi}^{(lead, CZ)}(s)|\simeq 0.4\,{\rm GeV}^2$ obtained with $\phi_{\pi}(x, \ov\mu\sim 1\,GeV)=\phi_{\pi}^{CZ}(x)$ and taking here the effective value of the coupling ${\ov\alpha}_s\simeq 0.4$. It is seen that there is a reasonable agreement. At the same time, using $\phi_{\pi}(x,\ov\mu)\simeq \phi^{asy}(x)$ one obtains the much smaller value $0.88\cdot |s\,F_{\pi}^{(lead, asy)}(s)|\simeq 0.13\,{\rm GeV}^2$.

Therefore, for the pion wave function $\phi_{\pi}(x, \mu)$ close to $\phi^{\rm asy}(x)$ the leading term (i.e. without power corrections) calculation predicts the cross section which is $\simeq 15$ times smaller than the data. It seems that at $s=W^2=10-16\,{\rm GeV}^2$ the power corrections can not cure so large difference.

Moreover, if power corrections were dominant numerically at these energies then the cross sections $\sigma_o(\pi^+\pi^-)$ and $\sigma_o(K^+K^-)$ will decay more like $\sim 1/W^{10}$, rather than $\sim 1/W^6$.\\

The SU(3)-symmetry breaking, $d\sigma(K^+K^-)\neq d\sigma(\pi^+\pi^-),$ originates not only from different meson couplings, $f_K\neq f_{\pi}$, but also from symmetry breaking effects in meson wave functions, $\phi_{K}(x)\neq \phi_{\pi}(x)$, see Fig.4. (Let us recall: {\it the\, heavier\, is\, quark\, the\, narrower\, is\, wave\, function}). These two effects tend to cancel each other when using for the K-meson the wave function $\phi^{CZZ}_ {K}(x_s,x_u)$ obtained in \cite{czz} from the QCD sum rules. So, instead of the naive original prediction $\simeq (f_K/f_{\pi})^4\simeq 2.3$ with $\phi_{K}(x)=\phi_{\pi}(x)$ in \cite{BL}, the prediction in \cite{Maurice} for this ratio is close to unity, and this agrees with the data from Belle \cite{Nakaz} :
\vspace*{-1mm}
\beq
\frac{\sigma_o (\gamma\gamma\ra K^+K^-)}{\sigma_o (\gamma\gamma\ra \pi^+\pi^-)}=\left \{
\begin{array}[c]{ll}\displaystyle (f_K/f_{\pi})^4\simeq 2.3 \scriptstyle &
\begin{array}[c]{l}\hspace*{-4.0 cm}\rm{ Brodsky,\, Lepage} \,\, {\rm\cite{BL}} \end{array}

\\ & \\ \hline & \\ \simeq 1.06 & \begin{array}[c]{l} \hspace*{-4.2 cm} \rm{Benayoun,\,Chernyak} \,\,{\rm\cite{Maurice}}\end{array} \\ & \\ \hline & \\
(0.89\pm 0.04\pm 0.15)\hspace*{3.0 cm}  {\rm Belle} \,\,\, {\rm\cite{Nakaz}}
\end{array} \right. \nonumber
\eeq
\vspace*{5
mm}

The leading terms in the cross sections for neutral particles are much smaller than for charged ones, see (1). For instance, it was obtained in \cite{Maurice} that the ratio $d\sigma^{(lead)}
(\pi^o\pi^o)/d\sigma^{(lead)}(\pi^+\pi^-)$ varies from $\simeq 0.07$ at $\cos \theta =0$ to $\simeq 0.04$ at $\cos\theta =0.6$, while the ratio
\beq
\frac{d\sigma (\ov{ K^o} K^o)^{(lead)}}{d\sigma (\pi^o\pi^o)^{(lead)}}\simeq 1.3\cdot(4/25)\simeq 0.2\,. \nonumber
\eeq
Besides,
\beq
\frac{\sigma_{o}^{(lead)}(K_S K_S)}{\sigma_{o}^{(lead)}(K^+K^-)}\simeq 0.005\,.\nonumber
\eeq

It is seen that the leading contribution to $\sigma_o(K_SK_S)$ is very small. This implies that it is not yet dominant at present energies $ W^2 < 16\,{\rm GeV}^2$. I.e., the amplitude $M(\gamma\gamma\ra K_S K_S)
= a(s,\theta)+b(s,\theta)$ is dominated at these energies by the non-leading term $b(s,\theta)\sim \varrho(
\theta)(s_o/s)^2$, while the formally leading term $a(s,\theta)\sim C_o f_{BC}(\theta)(s_o/s)$ has so small coefficient $|C_o|\ll 1$, that $|b(s,\theta)| > |a(s,\theta)|$ at, say, $s=W^2 < 12\,{\rm GeV}^2$.

Therefore, it has no much meaning to compare the leading term prediction of \cite{Maurice}, i.e.
\beq
\frac{d\sigma(K_S K_S)}{d \cos\theta}\sim |f_{BC}(\theta)|^2/W^{6}\quad  {\rm at}\,\, s\ra \infty
\nonumber
\eeq
for the energy and angular dependence of $d\sigma(K_S K_S)$ with the data from Belle \cite{Chen} at $W^2<15\,GeV^2$. Really, the only QCD prediction for, say, $6\,{\rm GeV}^2 < W^2 < 12\,{\rm GeV}^2$ is the expected energy dependence
\beq
\frac{d\sigma(K_S K_S)}{d\cos\theta} \sim \frac{|b(s,\theta)|^2}{s}\sim
\frac{|\varrho(\theta)|^2}{W^{10}}\,, \nonumber
\eeq
while the angular dependence $|\varrho(\theta)|^2$ and the absolute normalization are unknown. This energy dependence agrees with the results from  Belle \cite{Chen}, see fig.6\,.\\

\begin{minipage}[c]{.5\textwidth}\includegraphics
[trim=0mm 0mm 0mm 0mm, width=0.9\textwidth,clip=true]{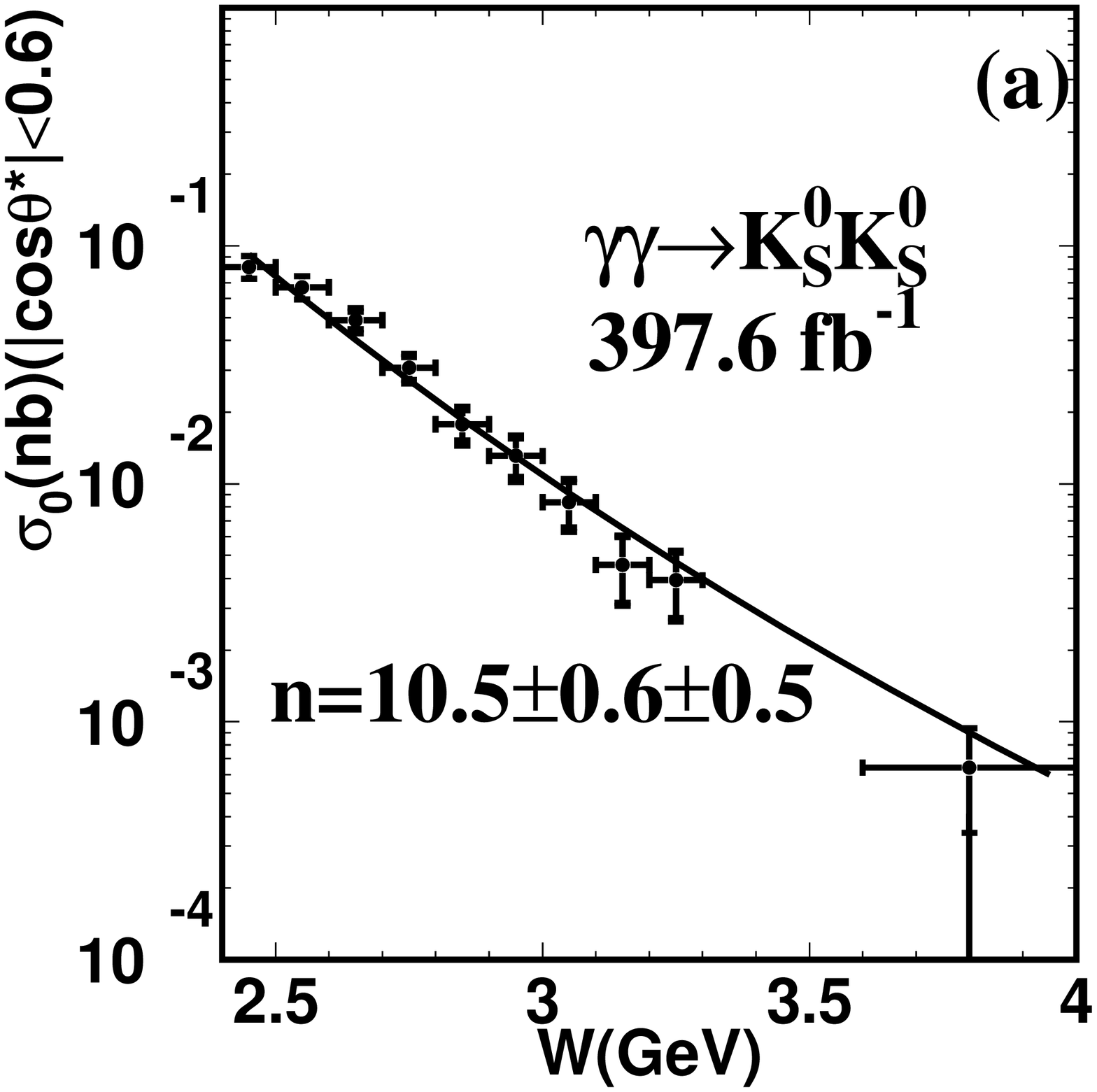}
\end{minipage}~
\begin{minipage}[c]{.5\textwidth}\includegraphics
[trim=0mm 0mm 0mm 0mm, width=0.9\textwidth,clip=true]{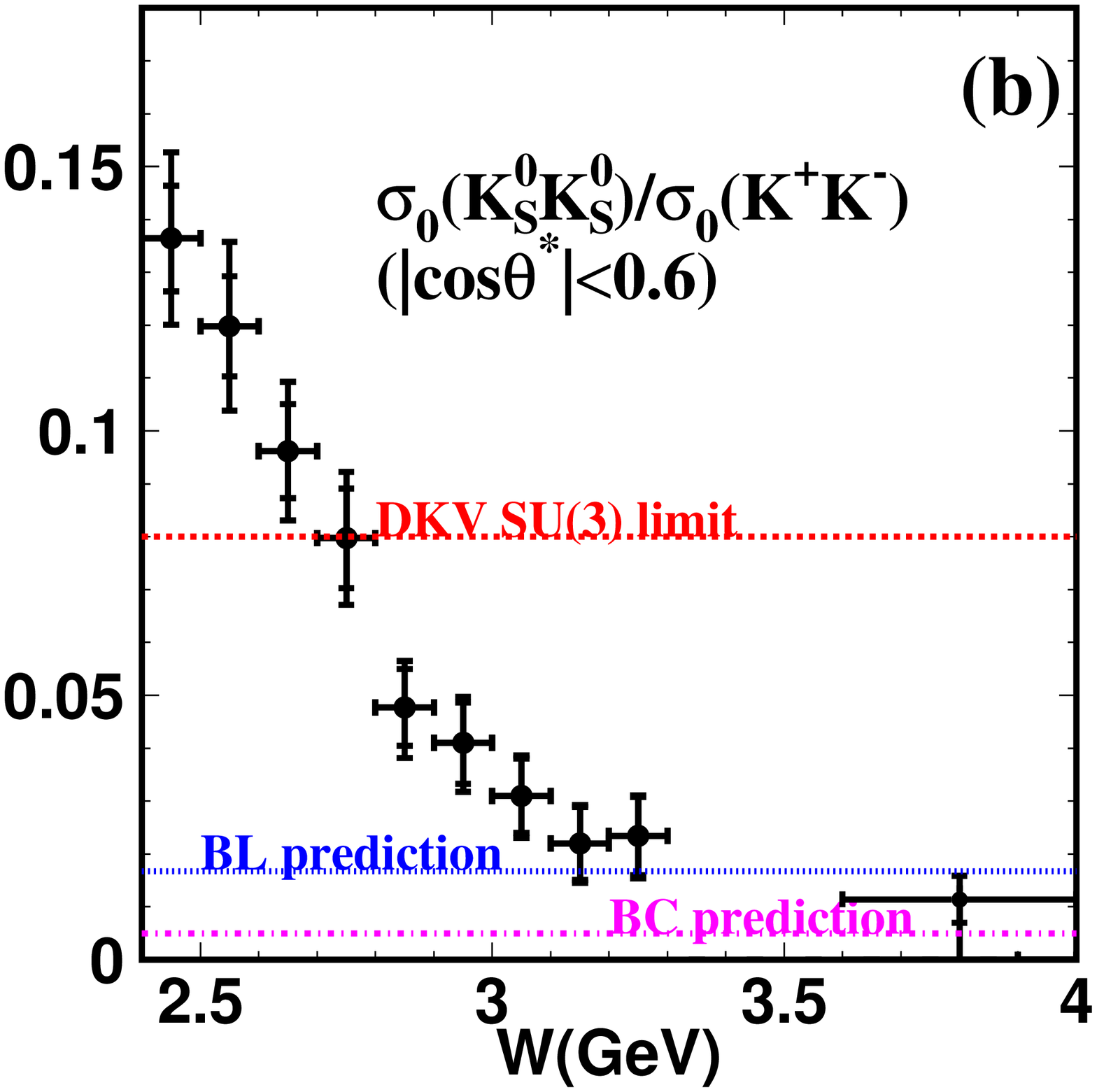}
\end{minipage}

\vspace{0.2 cm}
\hspace*{3mm} Fig. 6 \,\quad (a)\,\,  The total cross section $\sigma_o(\gamma\gamma\ra K_SK_S)$ in the c.m. angular region \\ $\hspace*{1.5cm}|\cos\theta|<0.6$\,\,\, \cite{Chen}. \, Here $n=(10.5\pm 0.6\pm 0.5)$ is the $W$-dependence $\sigma_{o}(W)\sim 1/W^{n}$\,.\\
\vspace{2mm}
\hspace*{8mm} Fig. 6\,\quad (b)\quad The ratio\, $\sigma_0(K_SK_S)/\sigma_0(K^+K^-)$ versus $W$\,\,\, \cite{Chen}\\
\hspace*{3mm} The dotted line DKV (Diehl-Kroll-Vogt) is the valence handbag model prediction in the $~SU(3)$ symmetry limit \cite{DKV}; the dashed BL and dash-dotted BC lines are the Brodsky-Lepage ( with $\,\phi_K(x)=\phi^{\rm asy}(x)\,$) and Benayoun-Chernyak ( with $\,\phi_K(x)=\phi^{CZZ}_K (x)\,$ from \cite{czz} ) leading term QCD predictions (for sufficiently large energy $W$).\\

The cross sections of other neutral particle productions were also measured by Belle Collaboration \cite{Ue-pi, Ue-eta, Ue-ee}, see Figs.7,8.\\

\begin{minipage}[c]{.5\textwidth}
\includegraphics[trim=0mm 0mm 0mm 0mm, width=0.75\textwidth,clip=true]{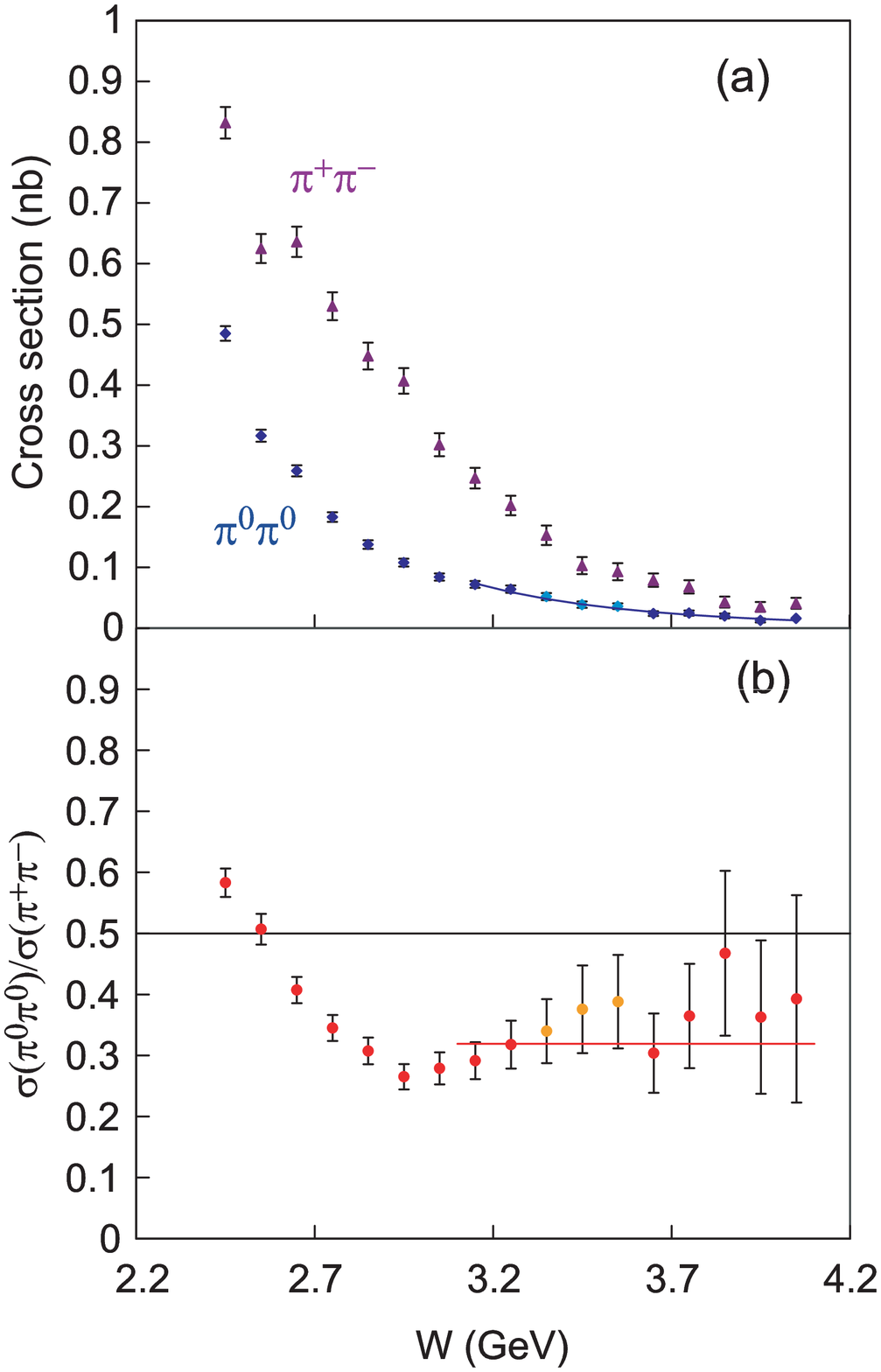}
\end{minipage}~
\begin{minipage}[c]{.5\textwidth}
\hspace*{-20mm} Fig. 7a\quad Cross sections $\sigma_{o}(\gamma\gamma\to \pi^0\pi^0)$ \\ and $\sigma_{o}(\gamma\gamma\to \pi^+\pi^-)$ for $|\cos \theta|<0.6$\,;

\hspace*{-20mm} Fig. 7b\quad Their ratio.  The lines are fits to the \\ results in the
energy region indicated.\\

\hspace*{-15mm} The QCD predictions for this range of energies:
\vspace*{-2mm}
\beq
\vspace*{-2mm}{\hspace*{-2cm}}\sigma(\pi^+\pi^-)\sim {\ov\alpha}_s^{\,2}/W^6\,,\nonumber
\eeq

{\hspace*{-2cm}} while the expected behavior of $\sigma(\pi^o\pi^o)$ if the higher twist\\
{\hspace*{-2cm}} terms are still dominant at $3<W<4\,GeV$ (and up to\\
{\hspace*{-2cm}} the odderon contribution, see below) is $~\sigma(\pi^o\pi^o)\sim 1/W^{10}$.
\vspace*{2mm}

{\hspace*{-1cm}} The handbag model prediction \cite{DKV}\,:
\vspace*{-2.5mm}
\beq
{\hspace*{-3cm}} R=\sigma(\pi^o\pi^o)/\sigma(\pi^+\pi^-)=0.5 \nonumber
\eeq
\end{minipage}

\newpage
\begin{minipage}[c]{.45\textwidth}\hspace*{-2mm}
\includegraphics[trim=0mm 0mm 0mm 0mm, width=1.3\textwidth,clip=true]{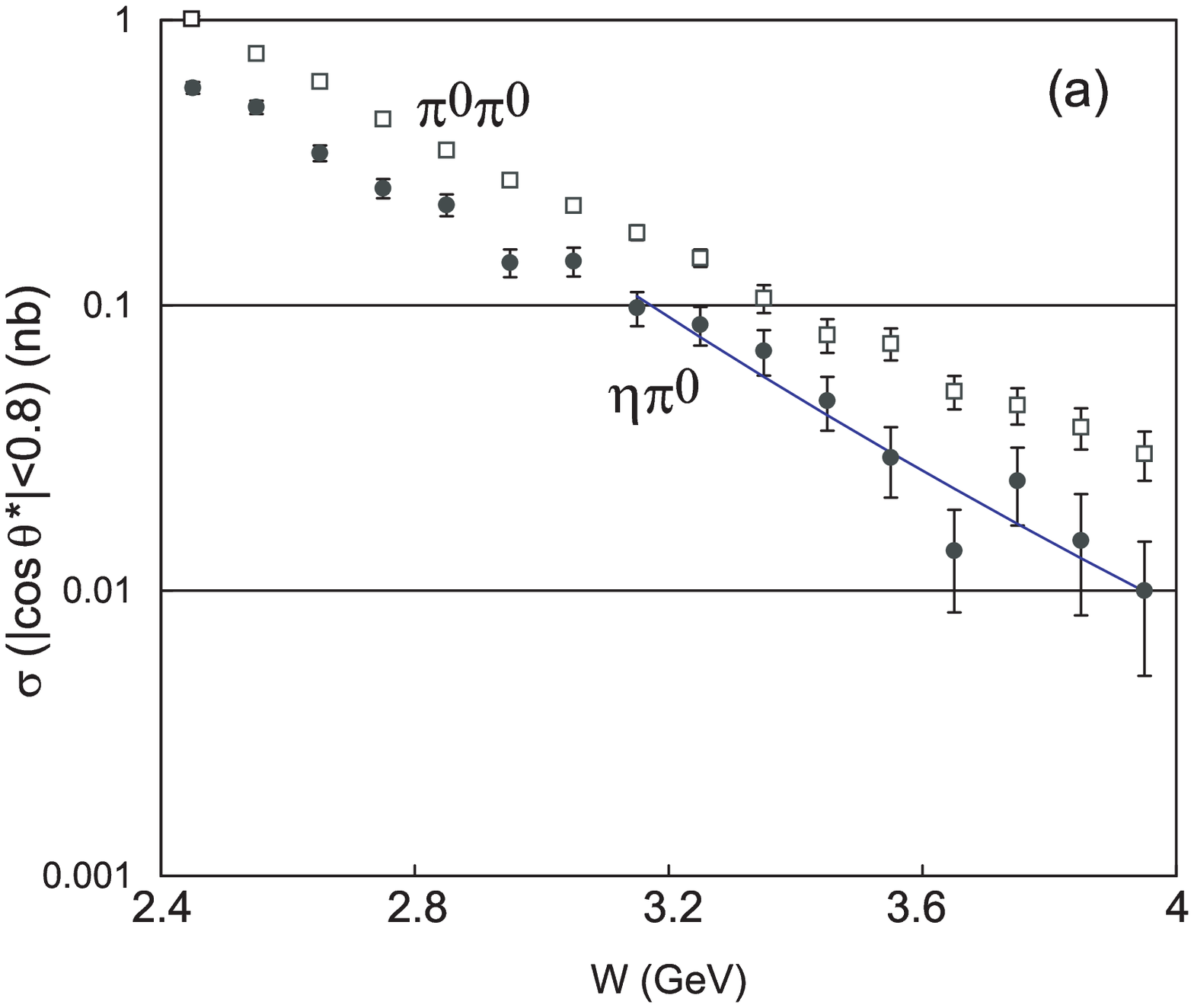}
\end{minipage}~
\begin{minipage}[c]{.55\textwidth}\hspace*{1mm}
\includegraphics[trim=0mm 0mm 0mm 0mm, width=0.85\textwidth,clip=true]{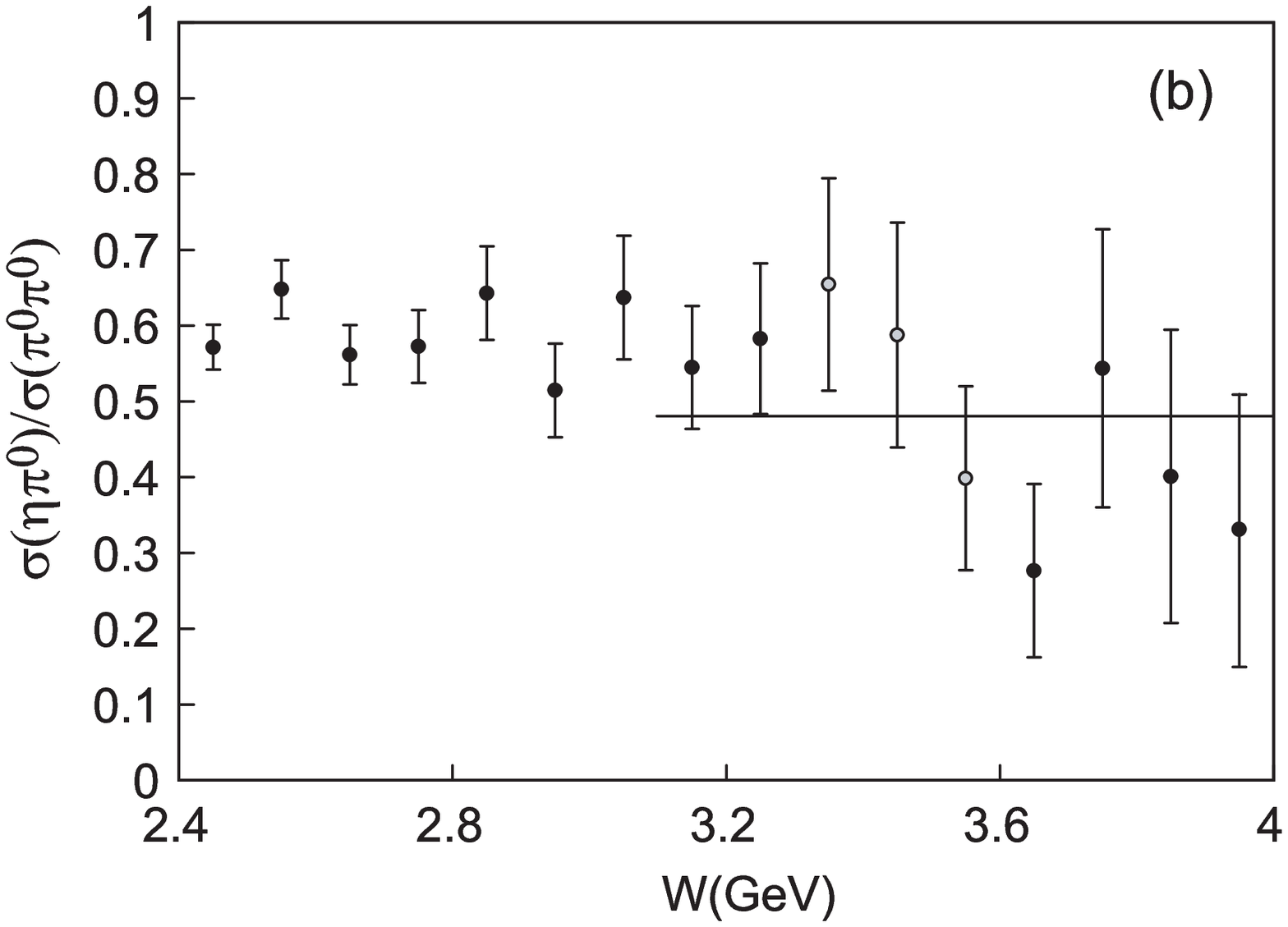}
\end{minipage}
\vspace{1mm}

Fig.8a \quad  $W$ - dependence of  cross sections $\sigma(\gamma\gamma\ra \pi^{o}\pi^{o})$ and $\sigma(\gamma\gamma\ra \eta\pi^o )\,,\, \,\, |\cos \theta|<0.8$.\\
{\hspace*{4cm}}The power low fit \,: $\sigma(\eta\pi^o)\sim (1/W)^n,\,\,~ n=(10.5\pm 1.2\pm 0.5)$\,\,\cite{Ue-eta}\\
\vspace{0.1cm}
{\hspace*{5mm}}Fig.8b \quad The ratio of cross sections $\sigma(\eta\pi^o)/\sigma(\pi^o\pi^o)$\,\,\cite{Ue-pi,Ue-eta}
\vspace*{4mm}

The energy dependencies of various cross sections measured and fitted by Bell Collaboration are collected
in the table.\\
\vspace*{1mm}

\begin{minipage}[c]{.99\textwidth}
\begin{center}
The value of \, "n" \, in $\sigma_{\rm tot}\sim 1/W^{n}$ in
various reactions \\ fitted in the $W$ and $|\cos\theta|$ ranges indicated
\end{center}
\hspace*{2.5mm}
\begin{tabular}{l|c|c|c|c|c|c} \hline \hline
Process & n - experiment & $W$ range (GeV) & $|\cos\theta|$ & n - QCD & n - handbag & Ref\\
\hline\hline
$\pi^+\pi^-$ & $7.9 \pm 0.4 \pm 1.5$ & $3.0 - 4.1$ & <0.6 & $\simeq 6$ & $\simeq 10$ & \cite{Nakaz} \\
\hline
$K^+K^-$  & $7.3 \pm 0.3 \pm 1.5$ & $3.0 - 4.1$ & <0.6 & $\simeq 6$ & $\simeq 10$ & \cite{Nakaz} \\
\hline
$K^0_S K^0_S$  & $10.5 \pm 0.6 \pm 0.5$ & 2.4 -- 4.0 & <0.6 & $\simeq 10$ & $\simeq 10$ &
\cite{Chen} \\
\hline
$\eta \pi^0$ & $ 10.5 \pm 1.2 \pm 0.5 $ & 3.1 -- 4.1 & <0.8 & $\simeq 10$ & $\simeq 10$ &
\cite{Ue-eta} \\
\hline
$\pi^0\pi^0$ & $\simeq 10$ & 2.5 -- 3.0 & <0.8 & $\simeq 10$  & $\simeq 10$ &
\cite{Ue-pi} \\
\hline
$\pi^0\pi^0$ & $8.0 \pm 0.5 \pm 0.4$\,\, & 3.1 -- 4.1 & <0.8 & $\simeq 10\,?$  & $\simeq 10$ &
\cite{Ue-pi}\\
\hline
$\pi^0\pi^0$ & $\simeq 10$\,\, & 2.5 -- 3.0 & <0.6 & $\simeq 10$  & $\simeq 10$ &
\cite{Ue-pi}\\
\hline
$\pi^0\pi^0$ & $6.9 \pm 0.6 \pm 0.7$\,\, & 3.1 -- 4.1 & <0.6 & $\simeq 10\,?$  & $\simeq 10$ &
\cite{Ue-pi}\\
\hline
$\eta\eta$ & $7.8\pm 0.6\pm 0.4$ & 2.4 -- 3.3 & <0.8 & $\simeq 10$ & $\simeq 10$ &
\cite{Ue-ee}\\
\hline\hline
\end{tabular}
\end{minipage}

\vspace{3mm}

\hspace*{3mm} The measured energy dependence of the $\pi^0\pi^0$ cross section is similar to $K_SK_S$ and
$\eta\pi^o$ cross sections at $6<W^2<9\, GeV^2$, but behaves "abnormally"\, in the energy interval $9<W^2<16\,GeV^2$.\\
\vspace{1mm}

\begin{minipage}[c]{.65\textwidth}\hspace*{-0cm}{\includegraphics
[trim=0mm 0mm 0mm 0mm, width=0.9\textwidth,clip=true]{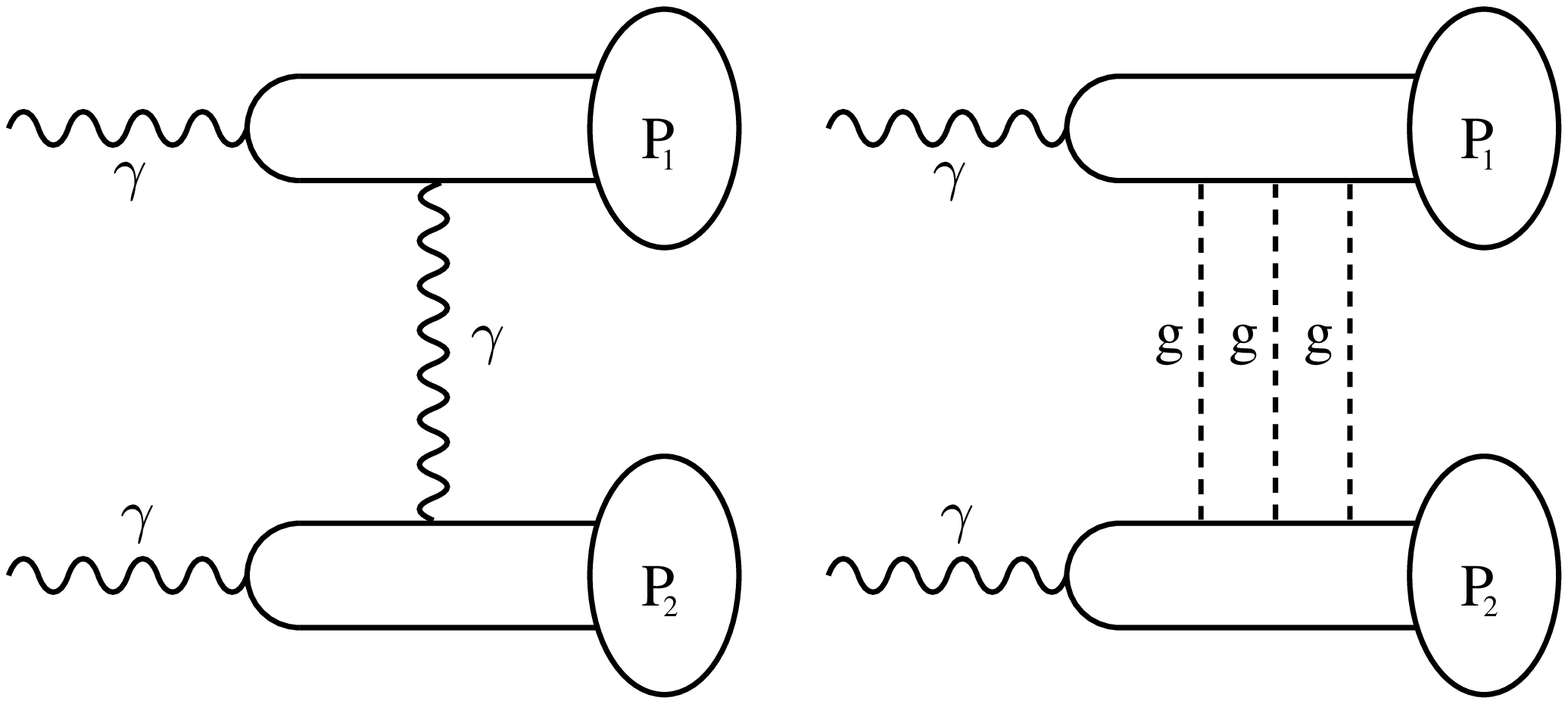}}
\end{minipage}
\begin{minipage}[c]{.35\textwidth}\hspace*{-1cm}
{Fig.9\quad  $P_1,\, P_2=\pi^o,\,\eta,\,\eta^\prime,\,\eta_c$ }\\

\hspace*{-2cm}{\hspace*{1cm} The additional hard contributions \\\hspace*{-0.6cm}for neutral
pseudoscalar mesons}
\end{minipage}
\vspace*{1mm}

\newpage
In attempt to understand this "abnormal" behavior of the $\pi^0\pi^0$ cross section we can recall that, unlike the $\sigma(\ov{K^o}K^o)$ cross section, there are additional contributions to the $\sigma(\pi^o
\pi^o)$ and $\sigma(\eta\pi^o)$ cross sections shown in Fig.9 (the odderon contribution becomes the leading one at sufficiently large energies and small fixed angles).

The contribution of the diagram with the photon exchange to the amplitudes $M_{\gamma}(\gamma\gamma
\to\pi^o\pi^o)$ is readily calculated and the helicity amplitudes look as
\beq
M_{\gamma}^{\pm\pm}=(4\pi\alpha)^2(e^2_u-e^2_d)^2\,\frac{2f^2_{\pi}}{s}\,\Bigl [\,\int_0^1\frac{\phi_{\pi}(x)}{x}\,\Bigr ]^2\,\Phi_{++}(\theta),\quad
\Phi_{++}(\theta)=\frac{2(1+z^2)}{(1-z^2)^2}\,, \nonumber
\eeq
\beq
M_{\gamma}^{\pm\mp}=(4\pi\alpha)^2(e^2_u-e^2_d)^2\,\frac{2f^2_{\pi}}{s}\,\Bigl [\,\int_0^1\frac{\phi_{\pi}(x)}{x}\,
\Bigr ]^2\,\Phi_{+-}(\theta),\quad \Phi_{+-}(\theta)=\frac{1+3z^2}{(1-z^2)^2}\,,
\quad z=\cos\theta\,.\nonumber
\eeq
As a result, using the pion wave function $\phi^{\rm CZ}_{\pi}(x)=30x(1-x)(2x-1)^2$ the ratio is (really,
this ratio is only weakly dependent on the pion wave function form)
\beq
\frac{\sigma_{\gamma}(\pi^o\pi^o,|\cos\theta|<0.8)}{\sigma(\pi^+\pi^-,|\cos\theta|<0.8)}\simeq 1\cdot\Bigl (\frac{\alpha}{{\ov\alpha}_s}\Bigr )^2\sim 0.5\cdot 10^{-3}\,,\nonumber
\eeq
so that this contribution is very small and does not help.\\

The odderon contribution in Fig.9 has been calculated in \cite{GI} and looks at $s\gg |t|\gg \mu^2_o$ as
\beq
M_{++}=M_{--}\simeq -2.5 M_{+-}=-2.5 M_{-+}\simeq \frac{sf^2_{\pi}}{t^2}\,\,4\pi\alpha\Bigl (\, 4\pi\ov{\alpha}_s\,\Bigr )^3\,\frac{5}{108\pi^2}\, I_{\pi\pi}\nonumber
\eeq
\beq
\hspace*{3cm} I_{\pi\pi}=\int_{-1}^{1}
d\xi_1\,\frac{\phi_{\pi}(\xi_1)}{(1-\xi^2_1)}\int_{-1}^{1}d\xi_2\,\frac{\phi_{\pi}(\xi_2)}{(1-\xi^2_2)}\,
T_{\pi\pi}(\xi_1,\xi_2),\hspace*{3cm} (4)\nonumber
\eeq
\beq
T_{\pi\pi}(\xi_1,\xi_2)=\ln \Big |\frac{\xi_1+\xi_2}{1-\xi_1}\Big |\ln \Big |\frac{\xi_1+\xi_2}{1+
\xi_2}\Big |+(\xi_1\ra -\xi_1),\quad\xi_1=x_1-x_2\,,\quad \xi_2=y_1-y_2\nonumber
\eeq
The numerical value of $I_{\pi\pi}$ in (4) is $I_{\pi\pi}\simeq 26.8$ \cite{GI} for $\phi_{\pi}(\xi)
=\phi^{\rm CZ}_{\pi}(\xi)$. Therefore, with $\phi_{\pi}(\xi)=\phi^{\rm CZ}_{\pi}(\xi)$\,:
\beq
\frac{d\sigma^{(3\,\rm gl)}}{dt}(\gamma\gamma\to\pi^o \pi^o)\simeq \frac{0.7\,nb\,GeV^6}{t^4}\Bigl (\,\frac{{\ov\alpha}^{\,2}_s}{0.1}\,\Bigr )^3, \nonumber
\eeq
and using here ${\ov\alpha}_s\simeq 0.3$\,:
\beq
\sigma^{(3\,\rm gl)}(\pi^o\pi^o,\,|\cos\theta|<0.8)\simeq \left\{\begin{array}{l l l}
\,\, 23\cdot 10^{-2}\,nb &  [\,\rm experiment: 30\cdot 10^{-2}\,nb\,]\,\, {\rm at}\,\, W=3\,GeV
\\
\,\,\,\, 9\cdot 10^{-2}\,nb & [\,\rm experiment:\,\,\, 8\cdot 10^{-2}\,nb\,] \,\, {\rm at}\,\, W=3.5\,GeV
\\
\,\,\,\,\, 4\cdot 10^{-2}\,nb & [\,\rm experiment:\,\,\, 3\cdot 10^{-2}\,nb\,] \,\, {\rm at}\,\, W=4\,GeV
\\
\end{array}\right. \nonumber
\eeq

Hence, according to these estimates with $\phi_{\pi}(\xi)=\phi^{\rm CZ}_{\pi}(\xi)$, the odderon
contribution is sufficiently large and may well be responsible for a change of the behavior of  $\sigma(\pi^o\pi^o)$ at $W>3$\,GeV. At the same time, the numerical value of $I_{\pi\pi}$ in (4) is $I_{\pi\pi}\simeq 7.4$ for $\phi_{\pi}(\xi)=\phi^{\rm asy}(\xi)$ and so the value of $\sigma^{(3\,\rm gl)}(\pi^o\pi^o)$ with $\phi_{\pi}(\xi)=\phi^{\rm asy}(\xi)$ will be $\simeq 13$ times smaller.

In the $SU(3)$ symmetry limit $\sigma^{(3\,\rm gl)}(\eta_8\pi^o)/\sigma^{(3\,\rm gl)}(\pi^o\pi^o)=2/3$. To estimate the effects of $SU(3)$ symmetry breaking we use the same model wave function of $\eta$ as those used in \cite{Ch3}, i.e. $|\eta\rangle=\cos\phi|n\rangle-\sin\phi|s\rangle,\, |n\rangle=|({\ov u}u+{\ov d}d)/
\sqrt 2\rangle, \,|s\rangle=|{\ov s}s\rangle$, and with taking into account the $SU(3)$ symmetry breaking effects distinguishing $|n\rangle$ and $|s\rangle$ wave functions.
Then, instead of $\phi^{\rm CZ}_{\pi}(\xi_1)$ in (4) one has to substitute
\beq
\phi^{\rm CZ}_{\pi}(\xi_1)\ra \Bigl [\,\frac{\cos\phi}{3}\,\phi^{\rm CZ}_\pi(\xi_1)+\frac{f_s}{f_\pi}\frac{\sqrt 2\sin\phi}{3}\,\phi^{\rm asy}(\xi_1)\,\Bigr ]\,.\nonumber
\eeq

Then, with $\phi\simeq 38^o$ and $f_s/f_\pi\simeq 1.3$, instead of $I_{\pi\pi}=26.8$ with $\phi_\pi(\xi)=\phi^{\rm CZ}_\pi(\xi)$, the corresponding integral will be $I_{\pi\eta}\simeq 11$ and $\sigma^{(3\,\rm gl)}(\pi^o\eta)/\sigma^{(3\,\rm gl)}(\pi^o\pi^o)\simeq 1/3$. It seems, this additional suppression may be a reason why the odderon contribution is still not seen clearly in $\sigma(\pi^o\eta)$ at $3<W<4\,GeV$ and $|\cos\theta|<0.8$. The prediction is that it will be seen at somewhat higher energies.

\begin{center}{\bf The handbag model}\end{center}

The hand-bag model \cite{DKV} is a definite application of the general idea which assumes that present day energies are insufficient for the leading terms QCD to be the main ones. Instead,  it is supposed that the soft nonperturbative contributions {\it dominate} the amplitudes. The handbag model realizes applications of this idea to a description of the large angle cross sections $d\sigma(\gamma \gamma\ra {\ov M}M)$.

As it is formulated in \cite{DKV}, the handbag model assumes that the above described hard QCD contributions really dominate at very high energies only, while the main contributions at present energies originate from the Fig.10a diagram.\\

\begin{minipage}{1.\textwidth}
\hspace*{3cm}\includegraphics[width=0.6\textwidth]{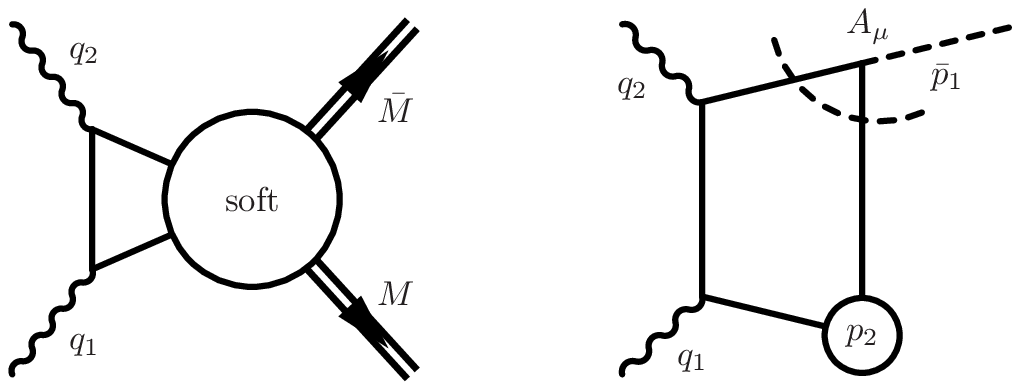}
\end{minipage}
\vspace{0.1mm}

Fig.10a\quad The overall picture of the handbag model contribution \cite{DKV}\\
\vspace{1mm}
\hspace*{3mm} Fig.10b\quad  The standard lowest order Feynman diagram for the light cone QCD sum rule \\
to calculate the soft valence handbag amplitude ${M_{\rm handbag}(\gamma\gamma\to \pi^+\pi^-)}$\,
\cite{Ch1,Ch2} ($A_\mu={\ov u}\gamma_\mu\gamma_5 d$)\\
\vspace*{1mm}

Here, {\it the two photons interact with the same quark only, and these two active ${\ov q}$ and $q$ quarks carry nearly the whole meson momenta, while the additional passive\, ${\ov q^\prime}$ and $q^\prime $ quarks are wee partons which are picked out from the vacuum by soft non-perturbative interactions} \cite{DKV}. Therefore, these soft form factors $R_{MM}(s)$ should be power suppressed in QCD at sufficiently large $s\,: R_{MM}(s) \leq 1/s^2\,,\, $ in comparison with the leading meson form factors, $F_{M}(s)\sim 1/s\,$. But, nevertheless, it is assumed that they are numerically dominant at present energies for both charged and neutral mesons.\\

The energy dependence and the absolute normalization of the handbag amplitude $M_{\rm handbag}(\gamma\gamma\\\ra {\ov M}_2 M_1)$ is not predicted in \cite{DKV} but fitted to the data.

As for the angular dependence, it was also not really predicted in \cite{DKV} in a model independent way. The reason is that a number of special approximate relations were used in \cite{DKV} at intermediate steps to calculate the angular dependence of the handbag amplitude. All these relations were valid, at best, for the leading term only. But it turned out finally that their would be leading term gives zero contribution to the amplitude, and the whole answer is due to next power corrections, $\sim \Lambda_{QCD}^2/s$, which were not under control in \cite{DKV}. The "result"\, $M_{\rm handbag}(\gamma\gamma\to M_2 M_1)\sim 1/\sin^{2}\theta$ for the handbag amplitude in \cite{DKV} is completely due to the one especially (and arbitrary) chosen definite power suppressed term in the amplitude while ignoring all other power corrections of the same order of smallness.

The authors were fully aware of this arbitrariness \cite{DKV}: {\it "We  must then at this stage consider our result $M_{\rm handbag}\sim 1/\sin^{2}\theta$ as a model or a partial calculation of the soft handbag contribution"}.

Hence, finally, the approach in \cite{DKV} predicts neither energy and angular dependencies nor the normalization of cross sections in a model independent way.

Therefore, what only remains are the specific predictions of the handbag model for the ratios of cross sections in the $SU(3)$ symmetry limit\,: {\it there is only one common valence handbag amplitude} $M^{\rm val}_{\rm handbag}$ (the soft non-valence handbag amplitudes are small, see below)\,:
\beq
M(\pi^+\pi^-)=M(\pi^o\pi^o)=M(K^+K^-)=\frac{5}{2}\, M(\ov{K^o} K^o)=\frac{5}{2}\, M(K_S K_S)=M^{\rm val}_{\rm handbag}\nonumber
\eeq
\beq
\frac{5}{\sqrt 3}\, M(\pi^o\eta_8)=\frac{5}{\sqrt 6}\, M(\pi^o\eta_o)=\frac{5}{3}\, M(\eta_8\eta_8)=\frac{5 }{2}\, M(\ov{K^o} K^o)=M^{\rm val}_{\rm handbag}\nonumber
\eeq

Due to these relations, the predictions of the handbag model for the ratios of cross sections in comparison with the data look as (the red numbers are specific predictions of the valence handbag model)\,:
\beq
\frac{\sigma (K^+K^-)}{\sigma (\pi^+\pi^-)}=1\,(0.89 \pm 0.04 \pm 0.15)_{\rm exp},\quad \frac{\sigma (\pi^o\pi^o)}{\sigma (\pi^+\pi^-)}={\color{red}{\frac{1}{2}}} \,(0.32\pm 0.03\pm 0.05)_{\rm exp} \nonumber
\eeq
\beq
\hspace*{-0.5cm}\frac{\sigma (K_S K_S)}{\sigma (K^+K^-)}={\color{red}{0.08}}\,\,(0.13\,\ra\,0.01,\, \rm{see\, Fig.6})_{\rm exp}, \quad\frac{\sigma (K_S K_S)}{\sigma(\pi^o\eta_8)}={\color{red}{\frac{2}{3}}}\,\,
(\,\sim 0.1\,)_{\rm exp}, \nonumber
\eeq
\beq
\frac{\sigma (\eta_8\eta_8)}{\sigma (\pi^o\pi^o)}= {\color{red}{0.36}}\, (0.37 \pm 0.04)_{\rm exp},\quad \frac{\sigma (\pi^o\eta_8)}{\sigma (\pi^o\pi^o)}={\color{red}{0.24}}\,(0.48 \pm 0.06)_{\rm exp}\nonumber
\eeq
\vspace*{0.5mm}

Recalling that the angular dependence of the handbag amplitude $M_{\rm handbag}\sim 1/\sin^2 \theta$ used in \cite{DKV} was a model form, it looks not so surprising that the explicit calculation of the valence handbag amplitude $M^{\rm val}_{\rm handbag}(W,\theta)$ in \cite{Ch1} (see also \cite{Ch2}) via the light cone QCD sum rules \cite{Braun},\cite{cz3} gave a different angular dependence, $M^{\rm val}_{\rm handbag}\sim {\rm const}$. These soft valence handbag contributions to the cross sections calculated explicitly from the light cone QCD sum rules in \cite{Ch1}, see Fig. 7b, are definite functions of the energy and scattering angle, and look as
$$\hspace*{3cm} d\sigma_{\rm handbag}(\gamma\gamma\to {\ov M}_2  M_1)/d\cos\theta \sim {\rm const}/W^{10}
\hspace*{4cm}(4)\nonumber $$ for all mesons, both charged and neutral.

Unfortunately, this angular behavior $d\sigma_{\rm handbag}(\gamma\gamma\to {\ov M}_2  M_1)/d\cos\theta\sim \rm const$ disagrees with all data which behave similar to $\sim 1/\sin^{4}\theta$, and the energy behavior  $\sigma_{\rm handbag}(\gamma\gamma\to {\ov M}_2  M_1)\sim 1/W^{10}$ disagrees with the data for charged mesons $\pi^+\pi^-$ and $K^+ K^-$, compatible with $\sim 1/W^6$.

This energy behavior $\sim 1/W^{10}$ in $(4)$ is as expected (up to Sudakov effects) in QCD for {\it soft} valence power corrections to the leading terms due to the {\it Feynman end-point mechanism} (but one should remember that there is also a number of {\it hard} valence power corrections in QCD with the same energy dependence $\sim 1/W^{10}$, but with possibly different angular dependencies).\\
\vspace*{-2mm}\hspace*{1.5cm}
\begin{minipage}[c]{0.6\textwidth}
\hspace*{-2cm}\includegraphics[trim=0mm 0mm 0mm 0mm, width=0.99\textwidth,clip=true]{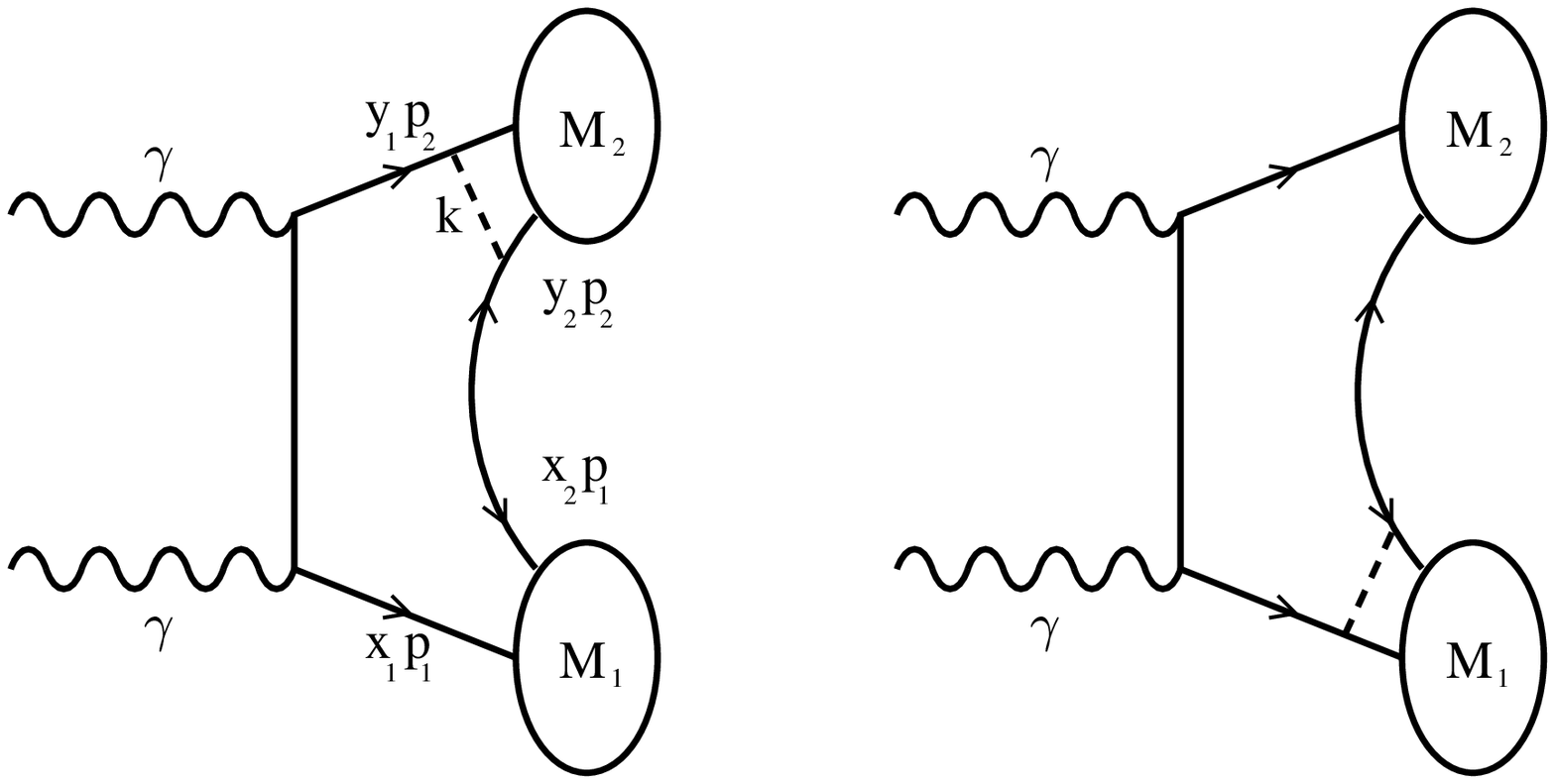}
\end{minipage}
\vspace*{0.5mm}
\begin{minipage}[c]{0.4\textwidth}
\hspace*{-3cm} Fig.11\quad The diagrams to calculate the energy \\\hspace*{-3cm} and angular dependencies of soft handbag\\ \hspace*{-3cm} amplitudes (the valence Feynman mechanism)
\end{minipage}

The end-point region contributions of only two diagrams shown in Fig.11 (the valence Feynman mechanism) are relevant for the standard valence handbag model as it is formulated in\,\,\,\cite{DKV}\,:\\
$\hspace*{1cm} x_1,\, y_1\ra 1,\,\,\,\quad k^2=x_2 y_2 S\sim \Lambda^2_{QCD}\,\,
\ra\,\, x_2\sim y_2\sim \delta=(\Lambda^2_{QCD}/S)^{1/2}=\Lambda_{QCD}/W\ll 1$.\\

The direct calculation of these diagrams in Fig.11 gives for the hard kernel
\beq
T^{(lead)}=const\, \frac{f^2_{\pi}}{S}\,\, \frac{(e_1 e_2)(x_1+y_1)+2 x_1 y_1}{x_1 x_2 y_1 y_2} \quad
\ra \quad const\, \frac{f^2_{\pi}}{S}\,\, \frac{(e_1 e_2)+1}{x_2 y_2}\,\hspace*{3cm} (5)\nonumber
\eeq
($e_1$ and $e_2$ are the photon polarization vectors), and for the soft end-point region contributions to the whole valence handbag amplitude
\beq
M_{\rm handbag}^{\rm val,\,\pm\pm}\sim \frac{f^2_{\pi}}{S}\Biggl [\int_0^{\delta}d x_2\Bigl (\frac{\phi_{\pi}(x)}{x_2}\sim const\Bigr )\Biggr ]^2\sim \frac{f^2_{\pi}}{S}\,\delta^2\sim \frac{f^2_{\pi}\Lambda^2_{QCD}}{S^2=W^4}\gg M_{\rm handbag}^{\rm val,\,\pm\mp}\,. \hspace*{2cm} (6)\nonumber
\eeq

Therefore, the angular and energy dependences of all cross sections resulting from these soft valence contributions (the Feynman mechanism) look as
$$\hspace*{4cm} d\sigma_{handbag}/d\cos\theta\sim |M|^2/W^2\sim 1/W^{10}\,. \hspace*{4.7cm} (7)\nonumber $$

The expressions (5)-(7) agree with the predictions of the light cone QCD sum rules \cite{Ch1} not only in the energy and angular dependencies but also in the photon helicities dependencies.\\

If we are interested only in the energy dependence of such soft end-point region contributions, there is
a simpler way to obtain it which does not require the direct calculation of Feynman diagrams. This can be done as follows.

1)\,\, There is the hard part of any such diagram and it is the amplitude of the annihilation of two photons into a pair of active near mass-shell quarks with each one carrying nearly the whole meson momenta, $A_{\rm hard}(\gamma\gamma\to {\ov q} q)$. From the dimensional reasons it is $A_{\rm hard}(\gamma\gamma\to {\ov q} q)\sim 1$.

2)\,\, All other parts of the Feynman diagrams are soft and, parametrically, depend on the scale $\Lambda_{QCD}$ only. So, the energy dependence of the soft valence end-point region contributions (i.e. the valence Feynman mechanism) looks here as follows ($\,\phi_2(x_1,x_2)\sim x_1 x_2\,$)
\beq
R^{(\rm v,\,2)}_{MM}(W)\sim M_2(\gamma\gamma\to {\ov M} M)\sim \int_0^{\delta}dx_2\,\phi_2(x)\int_0^{\delta}dy_2\,\phi_2(y)\Bigl [A_{\rm hard}(\gamma\gamma\to {\ov q} q)\sim 1 \Bigr ]\sim \nonumber
\eeq
\beq
\sim \Bigl [\,\int_0^{\delta}dx_2\, x_2\int_0^{\delta}dy_2\, y_2\,\Bigr ]^{\rm n_{\rm wee}=1}\sim \Bigl (\,\delta^4\,\Bigr )^{\rm n_{\rm wee}=1}\sim\Lambda_{QCD}^4/W^4\,. \nonumber
\eeq

This method of obtaining the energy dependence of soft end-point contributions will be used below to calculate the energy dependence of soft non-valence handbag form factors originating from the 4-particle components of meson wave functions.\\

The updated predictions of the handbag model for the $\gamma\gamma\to {\ov M_2}M_1$ cross sections were given in the next paper \cite{DK}. In comparison with the original paper \cite{DKV}, the main new element in \cite{DK} is that (in the $SU(3)$ symmetry limit used in \cite{DK} and \cite{DKV}) the sizeable soft non-valence form factor $R^{\,\rm nv}_{\ov M M}(s)$ is introduced now, in addition to the soft valence one $R^{\,\rm v}_{\ov M M}(s)$ (the soft non-valence contributions were neglected in \cite{DKV}). Both functions, $R^{\,\rm v}_{{\ov M}M}(s)$ and $R^{\,\rm nv}_{{\ov M}M}(s)$, are parameterized then in arbitrary forms with a number of free parameters which are fitted in \cite{DK} to the data.
\footnote{\,
The form factors $R^{\,\rm u}_{2\pi}(\rm s)$ and $R^{\,\rm s}_{2\pi}(s)$ used in \cite{DK} are connected with those from \cite{Ch3} as: $R^{\,\rm u}_{2\pi}(s)=R^{\,\rm v}_{2\pi}(s)+R^{\,\rm nv}_{2\pi}(s)$,\,\, $R^{\,\rm s}_{2\pi}(s)=R^{\,\rm nv}_{2\pi}(s)$\,.\\
}
\\

As for the soft valence contributions to the cross sections and the soft valence form factors $R^{\,\rm v}_{{\ov M}M}(s)$, these were estimated numerically in \cite{Ch1} via the standard light cone QCD sum rules and were found much smaller numerically (and with the expected suppressed power behaviour $R^{\,\rm v}_{{\ov M}M}(s)\sim 1/s^2$) than the values fitted to data in \cite{DKV} and \cite{DK}.\\

As for the soft non-valence contributions, the two types of such contributions are presented in Fig.12\, \cite{Ch3}.\\

\begin{minipage}[c]{1.1\textwidth}\hspace*{0.5cm}\includegraphics
[trim=0mm 0mm 0mm 0mm, width=0.75\textwidth, clip=true]{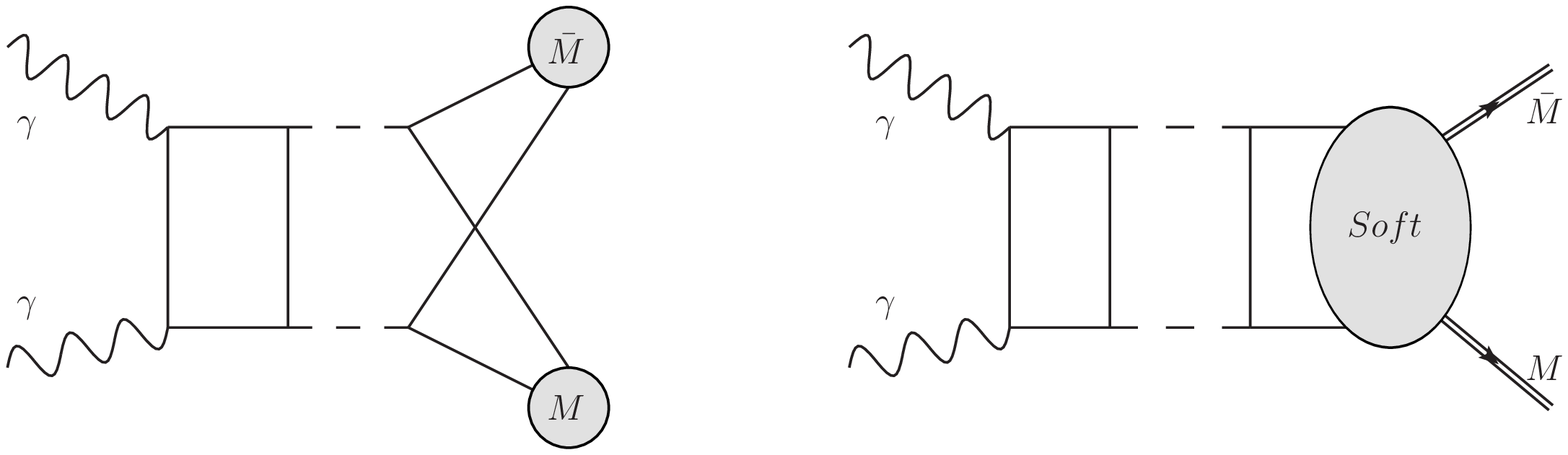}
\end{minipage}~
\vspace*{3mm}

\hspace*{2.5mm} Fig.12a \quad The leading power hard non-valence one-loop correction. \\
\vspace*{2mm}
\hspace*{7mm} Fig.12b \quad The leading contribution to the soft non-valence handbag form factor $R^{\,\rm nv}_{{\ov M}M}(s)$.\\
\vspace*{1mm}
\hspace*{15mm} The solid and dashed lines represent quarks and gluons.\\
\vspace*{3mm}

It is worth noting that both non-valence contributions in Fig.12 are $SU(3)$-flavor singlets in the $SU(3)$-symmetry limit. So, they contribute equally to the amplitudes $\pi^+\pi^-,\,
\pi^0\pi^0,\,K^+K^-,\\ \ov{K^0}K^0$ and $\eta_8\eta_8$, and don't contribute to $\eta_8\pi^0$.\\

The diagrams in Fig.12a constitute a small subset of all one-loop corrections to the leading power contributions from the Born diagrams like those shown in Fig.1\,. If these leading power one-loop non-valence corrections to the Born contributions were really significant, this will contradict then the data on $K_SK_S$,\, see Fig.6\,.

In particular, this hard non-valence one-loop correction was calculated, among all others, in \cite{DN}. Its contribution into the cross section $\sigma(\gamma\gamma\to K^+K^-)$ (integrated over $|\cos\theta|<0.6$, and with $\phi_K(x)=\phi^{\rm asy}(x)$ ) is \cite{Duplan}\,:
\beq
\frac{\delta\sigma^{\rm nv}}{\sigma} \simeq -\frac{{\ov\alpha}_s}{3\pi}\simeq -3\%\,,\nonumber
\eeq
i.e., its contribution into the amplitude is: $\delta{\ov A}^{\,\rm nv}/\,{\ov A}(K^+K^-)\simeq -1.5\%$.

The leading term amplitude $|{\ov A}(K_S K_S)|\simeq 0.15\, |{\ov A}(K^+K^-)|$,\, see Fig.6. Hence, the rough estimate of this non-valence one loop correction to the ${\ov A}(K_S K_S)$ amplitude is\,: \\ $|\delta{\ov A}^{\,\rm nv}/\,{\ov A}(K_S K_S)|\simeq 10\%$.\\

As for the soft non-valence handbag form factor $R^{\,\rm nv}_{{\ov M}M}(s)$, it seems sufficient to say that the leading contribution to it originates first from the Fig.12b two-loop correction ((without large logarithms) \cite{Ch3}, so that the estimate looks as\,:
\beq
\hspace*{3cm}\frac{R^{(\rm nv,\,2)}_{M M}(s)}{R^{\,\rm v}_{M M}(s)}\sim\Biggl (\frac{{\ov\alpha}_s}{\pi}\Biggr )^2\sim 0.01\,.\hspace*{4cm} (8)\nonumber
\eeq
\vspace*{1mm}

Besides, there are also soft non-valence contributions from the 4-quark components of the meson wave functions. For instance, the typical contribution of the pion 4-quark components, $|\pi^+\rangle_4\sim |
({\ov s}s+{\ov u}u+{\ov d}d)\,{\ov d}u\rangle$, is shown in Fig.13. The energy dependence of such contributions can be obtained the same way as for the diagram in Fig.11

\begin{minipage}[c]{.5\textwidth}
\includegraphics[trim=0mm 0mm 0mm 0mm, width=0.75\textwidth,clip=true]{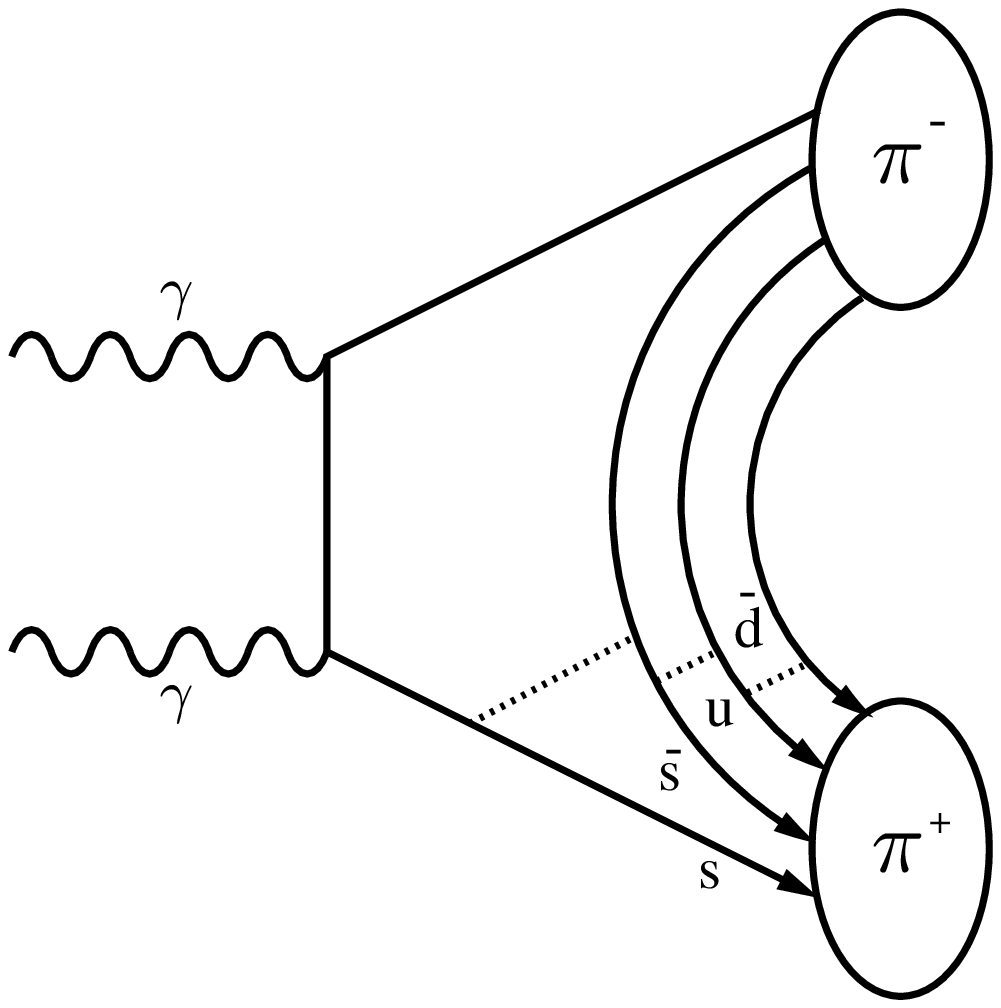}
\end{minipage}
\begin{minipage}[c]{0.4\textwidth}\vspace*{-5mm}

Fig.13\quad The additional contribution $R^{(\rm nv,\,4)}_{2\pi}(W)$ to the soft non-valence handbag form factor. Here, one strange quark in $\pi^+$ and $\pi^-$ carries nearly the whole pion momentum, while three other quarks are wee partons ($\,\phi_4(x)\sim x_1 x_2 x_3 x_4,\,\, \, x_1\ra 1,\,\,\, 0<x_{2,3,4}<\delta\sim \Lambda_{QCD}/W$)
\end{minipage}
\vspace*{1mm}
\beq
R^{(\rm nv,\,4)}_{2\pi}(W)\sim M_4(\gamma\gamma\to\pi^+ \pi^-)\sim\int_0^{\delta}d x_2 d x_3 d x_4\,\phi_4(x)\int_0^{\delta}d y_2 d y_3 d y_4\,\phi_4(y)\Bigl [A_{\rm hard}(\gamma\gamma\to{\ov s}s)\sim 1 \Bigr ] \nonumber
\eeq
\beq
\sim \Bigl [\,\int_0^{\delta}dx\, x\int_0^{\delta}dy\, y\,\Bigr ]^{\rm n_{\rm wee}=3}\sim \Bigl (\,\delta^{4}\,\Bigr )^{n_{\rm wee}=3}\sim\Lambda_{QCD}^{12}/W^{12}\,, \nonumber
\eeq
\beq
\hspace*{3.5cm}\frac{R^{(\,\rm nv,\,4)}_{2\pi}(W)}{R^{\,\rm v}_{2\pi}(W)}\sim \Bigl (\frac{1\,GeV^2}{W^2}\Bigr )^{4}\sim 1.5\cdot 10^{-4}\,\, \,\,{\rm at}\,\,s=W^2=9\,GeV^2\,.\, 
\hspace*{1.5cm} (9)\nonumber
\eeq

Clearly, so small soft non-valence contributions, $R^{(\rm nv,\,2)}_{M M}/R^{\,\rm v}_{M M}
\sim 10^{-2}$ and $R^{(\rm nv,\,4)}_{M M}/R^{\,\rm v}_{M M}\sim 10^{-4}$ at $W^2\simeq 9\,GeV^2$, can be safely neglected and will not help.\\
\vspace*{1mm}

In comparison,
\beq
\frac{|R^{\,\rm nv}_{2\pi}(W^2=9\, GeV^2)|}{|R^{\,\rm v}_{2\pi}(W^2=9\, GeV^2)|}\simeq 0.3 \nonumber
\eeq
was used in \cite{DK} to fit the data.\\
\vspace*{5mm}

\hspace*{5cm} {\bf The cross sections} {\boldmath{$\gamma\gamma\ra V_1^o V_2^o$}}\\
\vspace*{1mm}

The contribution of diagrams like Fig.1 to the amplitude $\gamma\gamma\to\rho^o\rho^o$ is also suppressed numerically in comparison with $\gamma\gamma\to\rho^+\rho^-$. But there is the additional hard leading twist one-loop contributions to $\gamma\gamma\to\rho^o_L \rho^o_L$, see Fig.14, and it looks at $s\gg |t|\gg \mu^2_o$ as \cite{cz4}\cite{GPS}\,:
\beq
M_{++}=M_{--}\simeq -2M_{+-}=-2M_{-+}\simeq i\,\frac{s f^2_{\rho}}{t^2}\,
(4\pi\alpha)\Bigl (4\pi{\ov\alpha}_s\Bigr )^2\frac{4}{9\pi}\,I_{\rho\rho}\nonumber
\eeq
\beq
I_{\rho\rho}=\int_{-1}^{1} d\xi_1\,\frac{\phi_{\rho}(\xi_1)}{(1-\xi^2_1)}\int_{-1}^{1}d\xi_2
\,\frac{\phi_{\rho}(\xi_2)}{(1-\xi^2_2)}\, T_{\rho\rho}(\xi_1,\xi_2)\nonumber
\eeq
\beq
T_{\rho\rho}(\xi_1,\xi_2)=\xi_1\xi_2 \ln \Big |\frac{\xi_1+\xi_2}{\xi_1-\xi_2}\Big |\,\,,
\quad \xi_1=x_1-x_2=2x_1-1,\quad \xi_2=y_1-y_2=2y_1-1\nonumber
\eeq
\vspace*{1mm}
\beq
\frac{d\sigma}{dt}(\gamma\gamma\to\rho^o_L \rho^o_L)\simeq \frac{0.9\,nb\,GeV^6}{t^4}, \quad
\sigma(\rho\rho,|\cos\theta|\leq 0.8)\simeq\frac{300\,nb\,GeV^6}{W^6}\nonumber
\eeq
(these numbers are for the $\rho$-meson wave function (11), see \cite{czz}, for $\phi_{\rho}\simeq\phi^{\rm asy}$ the cross section is $\simeq 6$ times smaller),
\beq
\hspace*{1cm}\sigma(\rho^o\rho^o,|\cos\theta|\leq 0.8)\simeq\left\{\begin{array}{l l l}
\,\, 41\cdot10^{-2}\,nb &   \,\, {\rm at}\quad W=3\,GeV \\
\,\, 16\cdot10^{-2}\,nb &   \,\, {\rm at}\quad W=3.5\,GeV \\
\,\,\,\,\, 7\cdot10^{-2}\,nb &   \,\, {\rm at}\quad W=4\,GeV \\
\end{array}\right. \hspace*{4cm} (10)\nonumber
\eeq

\begin{minipage}[c]{0.65\textwidth}
\includegraphics
[trim=0mm 0mm 0mm 0mm, width=0.85\textwidth, clip=true]{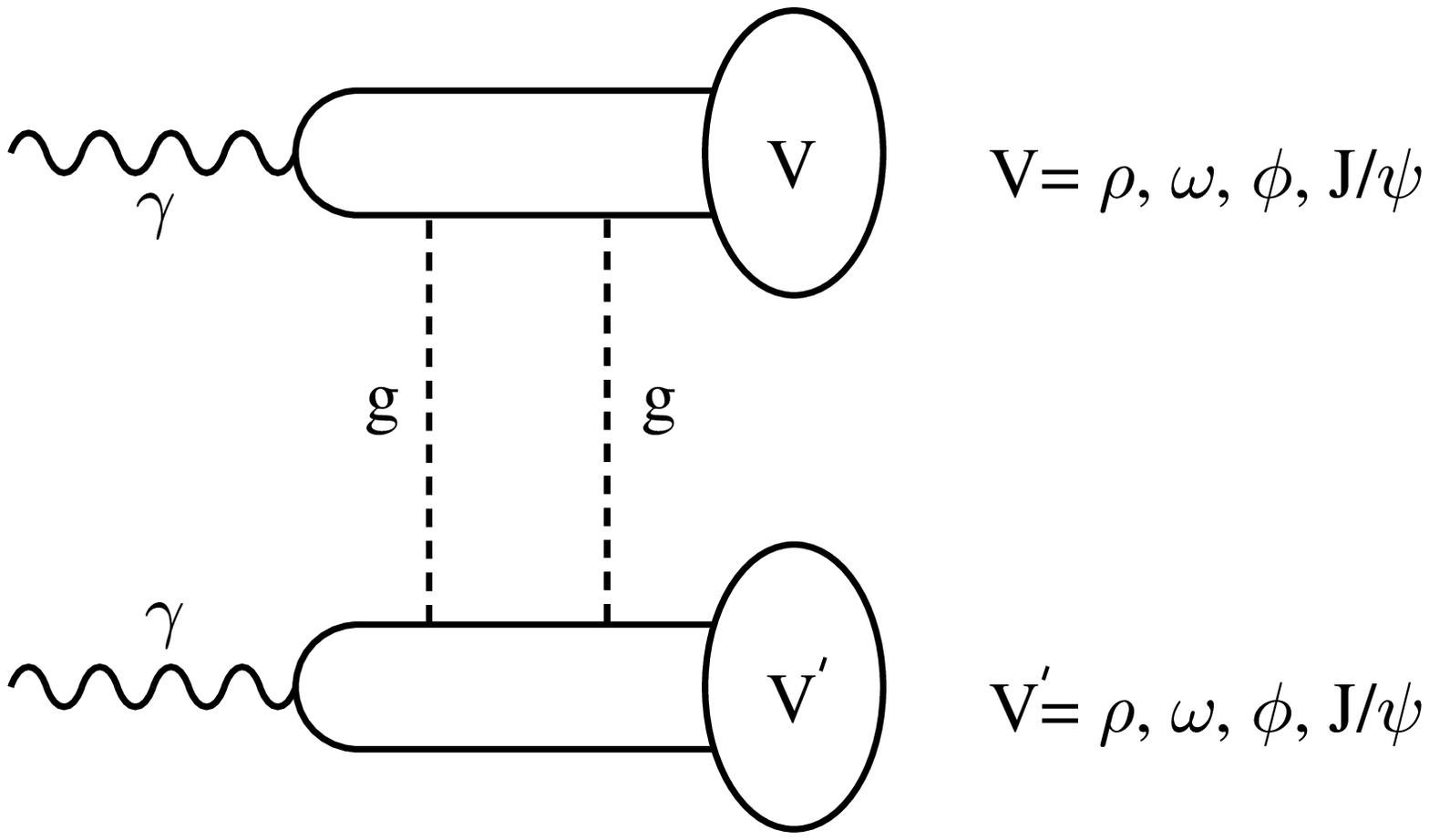}
\end{minipage}
\begin{minipage}[c]{0.35\textwidth \hspace*{-5mm}}
\hspace*{13mm} Fig.14\\ The additional hard one-loop contribution to the amplitude $\gamma\gamma\ra V_1^o V_2^o$. Becomes dominant at high energies and small fixed angles for longitudinally polarized vector mesons.
\end{minipage}
\vspace*{5mm}

\begin{minipage}[c]{0.45\textwidth}
\includegraphics [trim=0mm 0mm 0mm 0mm, width=0.99\textwidth,clip=true]{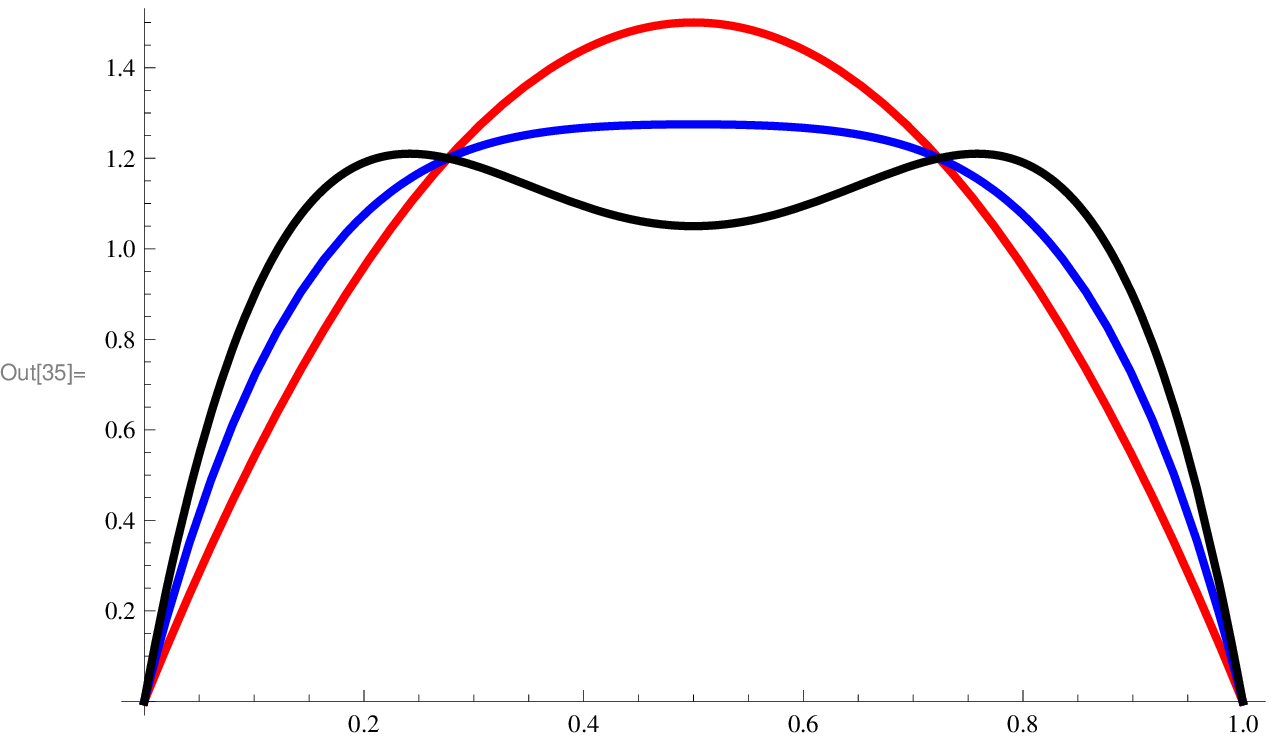}
\end{minipage}
\begin{minipage}[c]{0.55\textwidth \hspace*{0.5cm}}

\hspace*{5mm} Fig.15\,\, The model wave functions \\ Black line\,:\,\,  the $\rho_L$-meson wave function\\
 Blue\, line\,:\,\,  the $\phi_L$-meson wave function\\
 Red line\,:\,\,  the asymptotic wave function
\end{minipage}
\vspace*{1mm}

The contributions to other cross sections from the Fig.14 diagram
\beq
\sigma_o(\rho^o\rho^o): \sigma_o(\rho^o\omega): \sigma_o(\omega\omega)\simeq 1:\frac{1}{5}:\frac{1}{80}\nonumber
\eeq
\vspace*{1mm}

The model leading twist $V_{L}=V_{\lambda=0}$ wave functions \cite{czz} are taken in the form, see Fig.15
\beq
\phi_{\rho}(\xi,\mu\sim 1\,GeV)\simeq \phi_{\omega}(\xi,\mu_o\sim 1\,GeV)\simeq \phi^{\rm asy}(\xi)
\Bigl (1+0.2\, C^{3/2}_2(\xi)\Bigr )=\phi^{\rm asy}(\xi)\Bigl (0.70+1.5\,\xi^2\Bigr )\nonumber
\eeq
\beq
\phi_{\phi}(\xi,\mu_o\sim 1\,GeV)\sim \phi^{\rm asy}(\xi)
\Bigl (1+0.1\, C^{3/2}_2(\xi)\Bigr )=\phi^{\rm asy}(\xi)\Bigl (0.85+0.75\,\xi^2\Bigr ),\hspace*{2cm} (11)\nonumber
\eeq
\beq
\phi^{\rm asy}(\xi)=\frac{3}{4}(1-\xi^2),\quad
f_{\rho}\simeq f_{\omega}\simeq 210\,MeV\,,\quad f_{\phi}\simeq 230\,MeV\,.\nonumber
\eeq
\vspace*{1mm}

At $W=4\,GeV\,,\,\, |\cos\theta|\leq 0.8\,\,:$
\beq
\sigma(\omega\phi)\sim 3\,\sigma(\omega\omega)\sim 1.5\,\sigma(\phi\phi),\quad
\sigma(\rho^o\rho^o)\simeq 70\cdot 10^{-3}\,nb \nonumber
\eeq
\beq
\sigma(\omega\omega)\simeq 1\cdot 10^{-3}\,nb\quad \sigma(\omega\phi)\sim 3\cdot 10^{-3}
\,nb,\quad
\sigma(\phi\phi)\sim 2\,\sigma(\omega\omega)\sim 2\cdot 10^{-3}\,nb\,. \nonumber
\eeq
\vspace*{1mm}

The cross sections $\sigma(\omega\phi),\, \sigma(\phi\phi)$ and $\sigma(\omega\omega)$ have been
measured recently by Belle Collaboration \cite{LSY}, see Figs.16-18.

\newpage
\begin{minipage}[c]{.6\textwidth}
\includegraphics [trim=0mm 0mm 0mm 0mm, width=0.9\textwidth,angle=-90,clip=true]{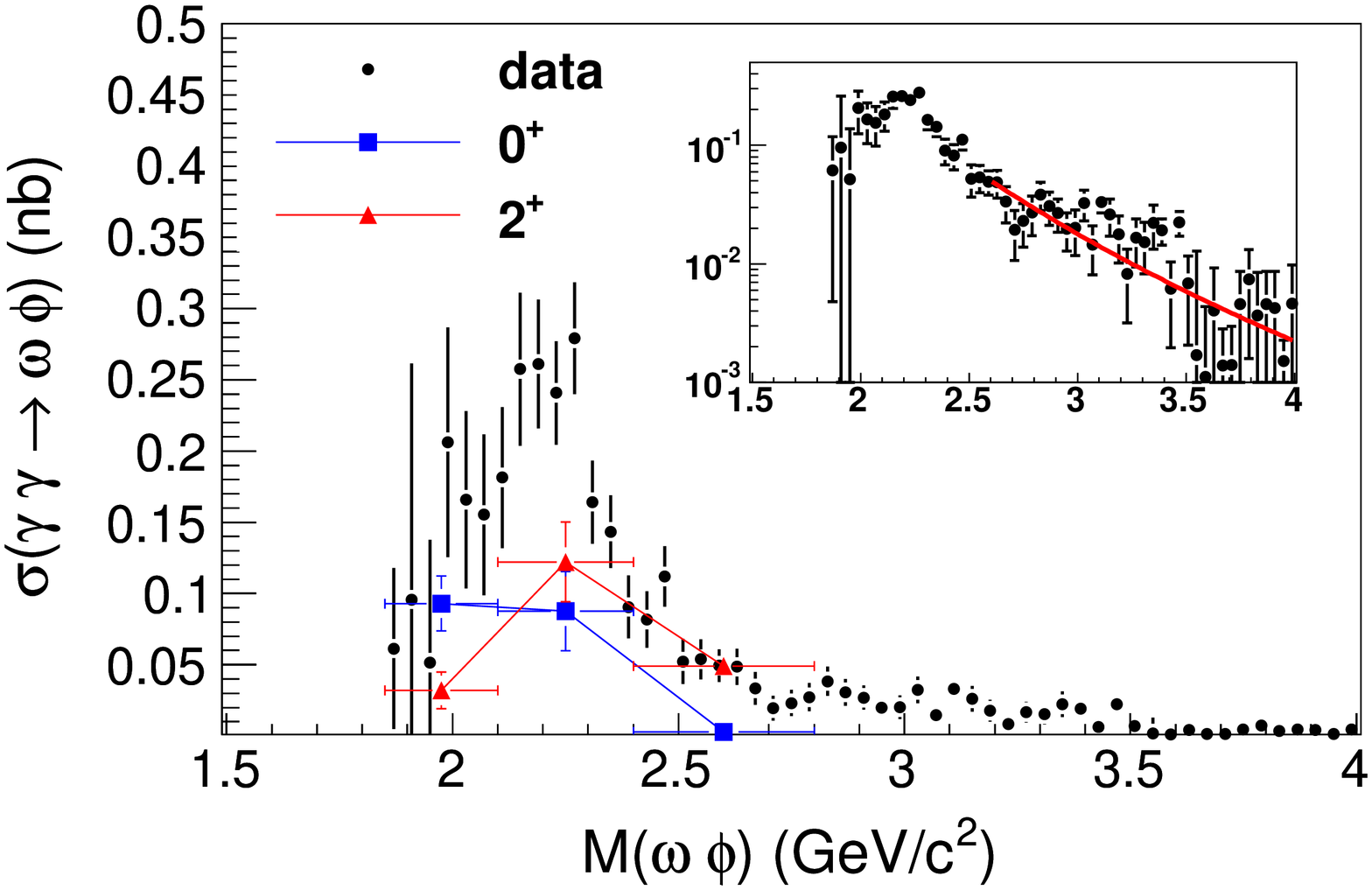}
\end{minipage}

\vspace{0.2 cm}

\hspace*{1cm} Fig.16 \quad   The total cross section $\sigma(\omega\phi)$ in
the c.m. angular region $|\cos\theta|< 0.8$ \cite{LSY}\\
\vspace*{2mm}
W-dependence\,:\, $\sigma_{\omega\phi}(W)\sim W^{-n},\,\, n=(7.2\pm 0.6_{stat}),\,\, \sigma(\omega\phi,\,W=4\,GeV)\simeq 2.5\cdot 10^{-3}\,nb$\\
\vspace*{2mm}
\hspace*{0.8cm} Theory:\,\, $\sigma(\omega\phi,\,|\cos\theta|\leq 0.8)\simeq 12\,nb\,GeV^6/W^6\,,\,\,\sigma(\omega\phi,\,W=4\,GeV)\simeq 3\cdot 10^{-3}\,nb$
\vspace*{5mm}

\begin{minipage}[c]{.6\textwidth}\includegraphics
[trim=0mm 0mm 0mm 0mm, width=0.9\textwidth,angle=-90,clip=true]{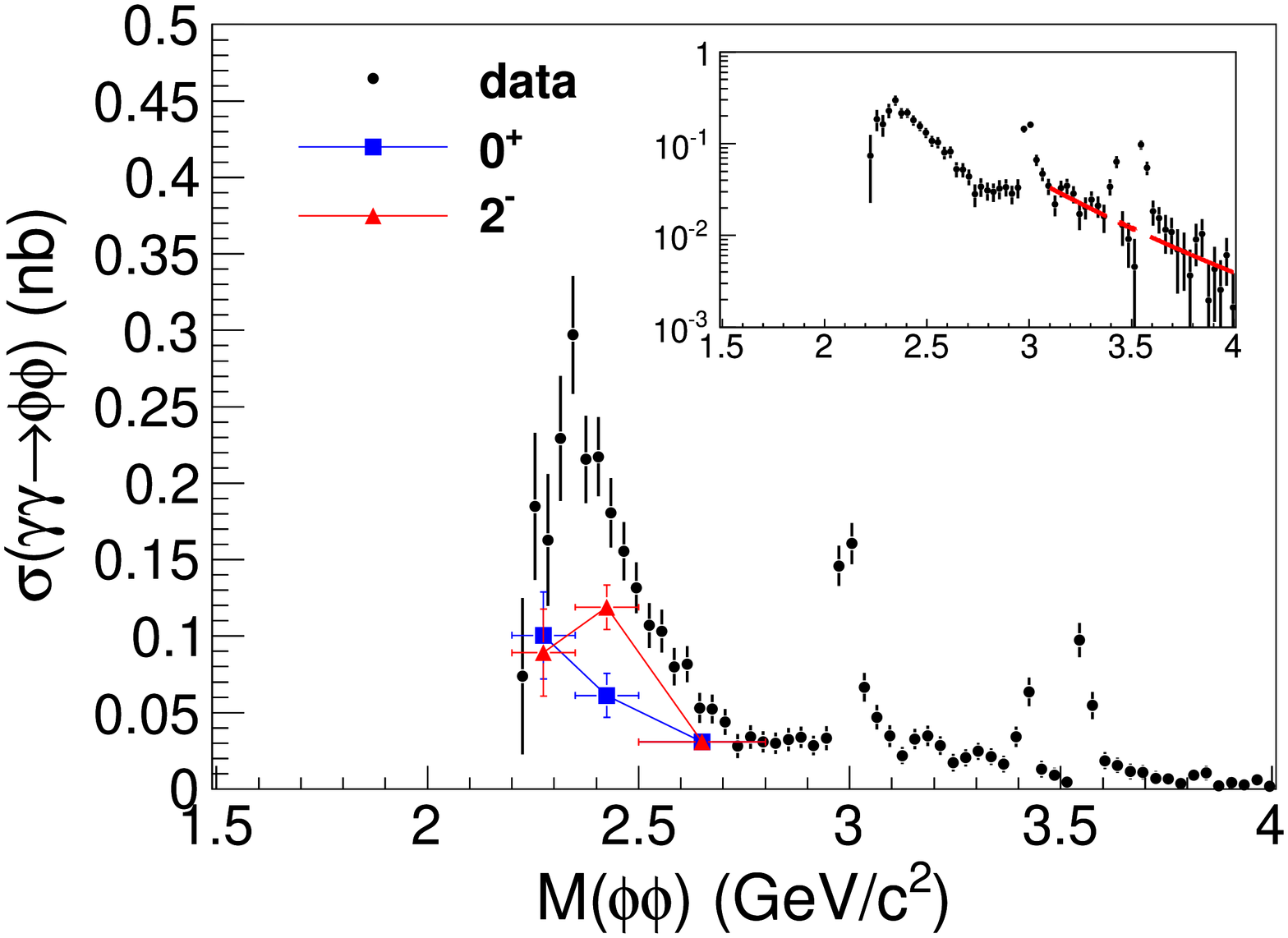}
\end{minipage}~
\vspace*{5mm}

\hspace*{1cm} Fig.17\quad  The total cross section $\sigma(\phi\phi)$ in the c.m. angular region $|\cos\theta|< 0.8$ \cite{LSY}.\\
\vspace*{2mm}
W-dependence:\, $\sigma_{\phi\phi}(W)\sim W^{-n},\,\, n=(9.1\pm 0.6_{stat}),\,\,\sigma(\phi\phi,\,W=4\,GeV)\simeq 3.5\cdot 10^{-3}\,nb$\\
\vspace*{2mm}
\hspace*{0.6cm} Theory:\, $\sigma(\phi\phi,\,|\cos\theta|\leq 0.8)\simeq 8
\,nb\,GeV^6/W^6,\,\,\sigma(\phi\phi, W=4\,GeV)\sim 2\cdot 10^{-3}\,nb$

\newpage
\begin{minipage}[c]{.6\textwidth}\includegraphics
[trim=0mm 0mm 0mm 0mm, width=0.9\textwidth,angle=-90,clip=true]{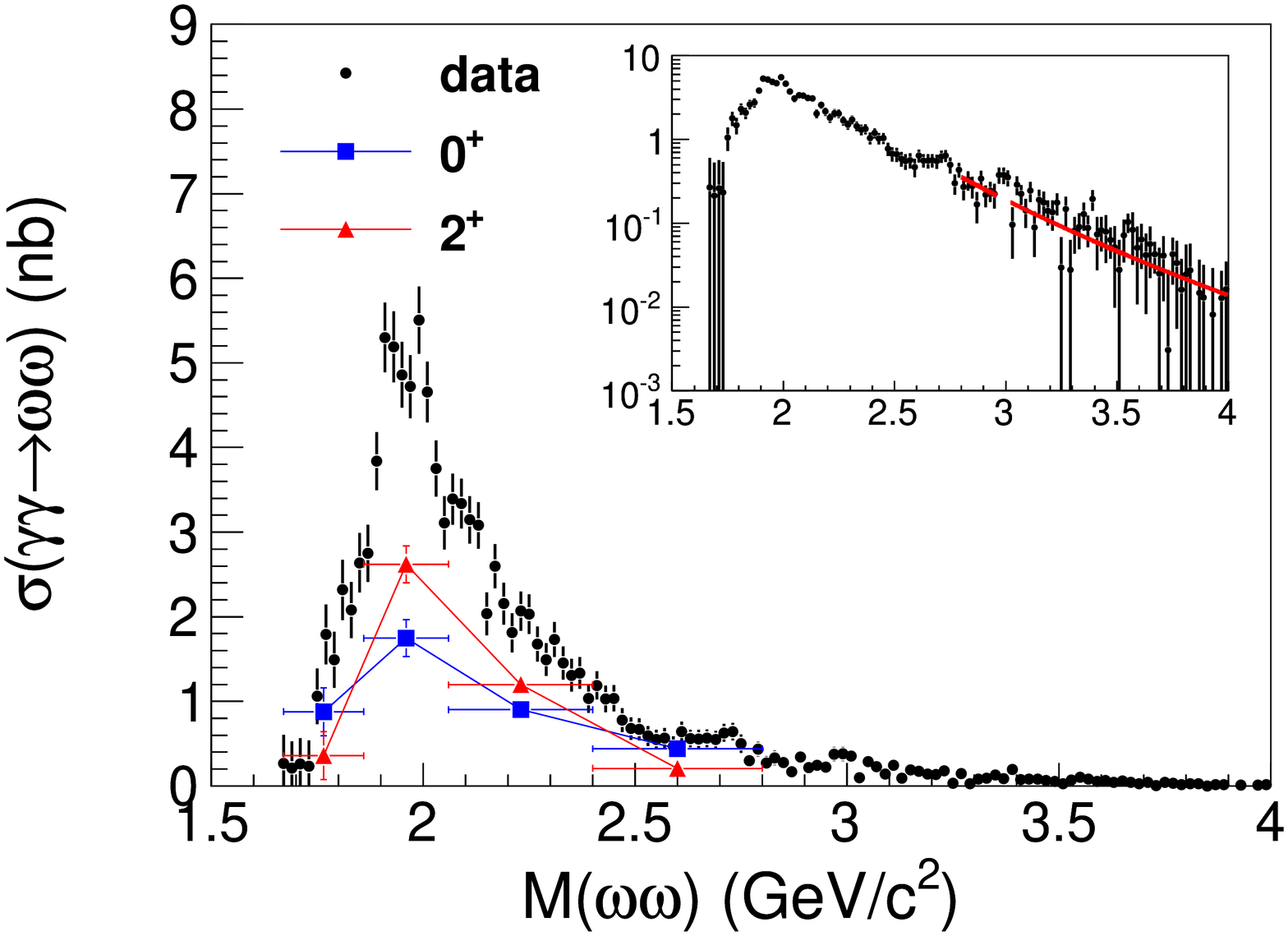}
\end{minipage}~
\vspace*{2mm}

\hspace*{1cm} Fig.18\quad  The total cross section $\sigma(\omega\omega)$
in the c.m. angular region $|\cos\theta|< 1.0$\\
\vspace*{2mm}
W-dependence:\, $\sigma_{\omega\omega}(W)\sim W^{-n},\,\, n=(8.4\pm 1.1_{stat}),\,\,\sigma(\omega\omega,\,W=4\,GeV)\simeq 15\cdot 10^{-3}\,nb$\\
\vspace*{2mm}
\hspace*{0.8cm} Theory:\,\,$\sigma(\omega\omega,\,|\cos\theta|\leq 0.8)\simeq 4\,nb\,GeV^6/W^6\,,\,\,
\sigma(\omega\omega,\,W=4\,GeV)\simeq 1\cdot 10^{-3}\,nb$
\vspace*{5mm}

It is seen that there is an agreement between the predicted and measured cross sections $\sigma
(\omega\phi,\,|\cos\theta|\leq 0.8)$ and $\sigma(\phi\phi,\,|\cos\theta|\leq 0.8)$ at $W=4\,GeV$.

The much larger measured cross section \cite{LSY}
\beq
\sigma(\omega\omega,\,|\cos\theta|<1)\gg \sigma(\phi\phi,\,|\cos\theta|\leq 0.8)\sim \sigma(\omega\phi,\,|\cos\theta|\leq 0.8) \nonumber
\eeq
looks natural at the first sight as one expects it is dominated by the forward region. But then it is strange that it decays so quickly, $\sigma(\omega\omega,\,|\cos\theta|<1)\sim 1/W^8$.

But the authors said \cite{LSY1}\,:

"We do not observe many events in $0.9<|\cos\theta|<1.0$ angle range for $\omega\omega$. Our measured cross section in the paper is for the whole angle range in Fig.2c in the paper \cite{LSY}.
\footnote{\,
Fig.18 here
}
If we require $|\cos\theta|<0.8$ for $\omega\omega$, the observed cross section dependence on the energy in the high energy region is similar. This is only our experimental observation. Due to very limited statistic, we can not measure cross section for the range of $0.9<|\cos\theta|<1$ in high energy region".

Also, it is said in the article \cite{LSY} that there is no detected events at $|\cos\theta|>0.8$ for $\omega\phi$ and $\phi\phi$.

But if, due to experimental restrictions, $\sigma(\omega\omega,\,|\cos\theta|<1)$ should be understood as
\\ $\sigma(\omega\omega,\,|\cos\theta| <0.8)$, why then is it so large\,:
\bq
\sigma(\omega\omega,\,|\cos\theta|<0.8)\sim 6\Biggl (\frac{\sigma(\omega
\phi,\,|\cos\theta|<0.8)+\sigma(\phi\phi,\,|\cos\theta|<0.8)}{2} \Biggr)\quad ?\nonumber
\eq
while the theory predicts
\bq
\sigma(\omega\omega)<\sigma(\phi\phi)<\sigma(\omega\phi)\,.\nonumber
\eq
\newpage
\begin{center} {\bf Short conclusions} \end{center}
\vspace*{3mm}

1)\,\, The leading term QCD predictions $d\sigma/d\cos\theta\sim {\ov\alpha}^{\,2}_s/(W^6\sin^{4}\theta)$ for {\it charged} mesons $\pi^{+}\pi^{-}$ and $K^{+}K^{-}$ agree reasonably well with data both in the energy and angular dependences at energies $W > 2.5-3\,GeV$ \cite{Nakaz}. The absolute values of cross sections agree reasonably well with data \cite{Nakaz} only for the wide pion (kaon) wave\, functions like $\phi_{\pi,K}^{\rm CZ}(x)$ \cite{cz2}, while the asymptotic wave functions $\phi_{\pi,K}(x)\simeq\phi^{\rm asy}(x)$ lead to cross sections one order smaller than data.\\
\vspace*{1mm}

2)\,\, In comparison with charged mesons, the QCD leading terms for neutral mesons are much smaller so that non-leading terms may be dominant at present energies $W<4\,GeV$ and in this case the energy dependence will be steeper, $\sigma({\ov {M^{o}}}M^{o})\sim 1/W^{10}$. This agrees with the data on $\sigma(K_S K_S)$ \cite{Chen} and $\sigma(\eta\pi^{o})$ \cite{Ue-eta},\, while $\sigma(\pi^{o}\pi^{o})$ is consistent with $\sim 1/W^{10}$ at $6<W^2<$ $9-10\,GeV^2$, but behaves "abnormally"\,, $\sim (1/W)^{7-8}$, at $10<W^2<16\, GeV^2$ \cite{Ue-pi}.

This can have natural explanation as, unlike $\sigma(K_S K_S)$, there is the additional odderon contribution to $\sigma(\pi^{o}\pi^{o})$, see Fig.9. With the pion wave function $\phi_{\pi}(x)=\phi^{\rm CZ}_{\pi}(x,\mu_o)$ it looks as
\beq
\sigma^{(3\,\rm gl)}(\pi^o\pi^o,\,|\cos\theta|<0.8)\simeq \Bigl (\,\frac{{\ov\alpha}^{\,2}_s}{0.1}\,\frac  {9\,{\rm GeV}^2}{W^2}\Bigr )^3\cdot 0.3\,nb \nonumber
\eeq
and may well be responsible for such an "abnormal" behavior of $\sigma(\pi^{o}\pi^{o})$ at $W^2>9-10\,{\rm GeV}^2$. At the same time, with $\phi_{\pi}(x)=\phi^{\rm asy}(x)$, the odderon contribution to the cross section $\pi^{o}\pi^{o}$ will be $\simeq 13$ times smaller.

The role of the odderon contribution to $\eta\pi^o$ is somewhat smaller in comparison with $\pi^{o}\pi^{o}$ and this may be a reason why it is still not seen clearly in $\sigma(\eta\pi^o)$ at $|\cos\theta|<0.8$ and $10<W^2<16\,GeV^2$. The prediction is that it will be seen here at somewhat higher energies.\\
\vspace*{1mm}

3)\,\,  In the Diehl-Kroll-Vogt approach \cite{DKV}, the handbag model gives definite model independent predictions neither for the energy nor the angular dependences of the cross sections $d\sigma
(\gamma\gamma\to {\ov M} M)$. The data are simply fitted with the model forms of amplitudes and a number of free parameters.\\
\vspace*{1mm}

4)\,\, The estimates of leading terms of the valence handbag amplitudes via the standard light cone QCD sum rules \cite{Ch1} or from the valence Feynman mechanism in QCD show that for all mesons, both charged and neutral, the soft handbag cross sections behave as
\beq
\frac{d\sigma_{\rm handbag}(\gamma\gamma\to M_2 M_1)}{d\cos\theta}\sim \frac{const}{W^{10}}\,.\nonumber
\eeq

This angular behavior disagrees with all data and the energy behavior disagrees with the data for $(\pi^+\pi^-)$ and $(K^+ K^-)$.\\
\vspace*{1mm}

5)\,\, In the $SU(3)$ limit, the relatively large non-valence soft handbag form factor, $R^{\,\rm nv}_{{\ov M}M}(s)\simeq\\ 0.3 R^{\,\rm v}_{{\ov M}M}(s)$ at $s=9\,GeV^2$, parametrized in a model form with a number of additional free parameters was used in the next paper of Diehl-Kroll \cite{DK} to improve fits to the data. Unfortunately, so large soft non-valence form factor contradicts the QCD estimates $R^{\,\rm nv}_{{\ov M}M}(s)\sim({\ov\alpha}_s/\pi)^2 R^{\,\rm v}_{{\ov M}M}(s)\sim 0.01 R^{\,\rm v}_{{\ov M}M}(s)$, see (8),(9).\\
\vspace*{1mm}

6)\,\, The very recent Belle data appeared on the cross sections of neutral vector mesons $\sigma(\gamma\gamma\to \omega\phi,\, \phi\phi,\, \omega\omega)$ \cite{LSY}. The values of the cross sections $\sigma(\gamma\gamma\to \omega\phi)$ and $\sigma(\gamma\gamma\to \phi\phi)$ at $W=4\,GeV$ and $|\cos\theta|<0.8$ are in a reasonable agreement with the theory predictions, but the fitted energy behavior at $2.7\,GeV<W<4\,GeV$ looks somewhat too steep (within large error bars at $3.5<W<4\,GeV$), in comparison with the leading term QCD predictions $\sigma(\gamma\gamma\to V_2^o V_1^o, |\cos\theta|<0.8)\sim {\ov\alpha}^{\,4}_s/W^6$.

But the value of the cross section $\sigma(\gamma\gamma\to \omega\omega)$ at $W=4\,GeV$ and $|\cos\theta|<0.8$ looks too large in comparison with the theory predictions.\\

I am grateful to organizers of this workshop Profs. Augustin E. Chen (NCU), Hsiang-nan Li (AS) and Sadaharu Uehara (KEK), and to Prof. Hai-Yang Cheng (AS) for a kind hospitality.\\

This work is supported in part by  Ministry of Education and Science of the Russian Federation and RFBR grant 12-02-00106-a.


\begin{thebibliography}{99}
\bibitem{cz1}
V.L. Chernyak, A.R. Zhitnitsky, JETP Lett. {\bf 25} (1977) 510\\
V.L. Chernyak, V.G. Serbo, A.R. Zhitnitsky, JETP Lett. {\bf 26} (1977) 594\\
V.L. Chernyak, A.R. Zhitnitsky, Sov. J. Nucl. Phys. {\bf 31} (1980) 544\\
V.L. Chernyak, V.G. Serbo, A.R. Zhitnitsky, Sov. J. Nucl. Phys. {\bf 31} (1980) 552
\bibitem{ER}
A.V. Efremov, A.V. Radyushkin, Phys. Lett. {\bf B94} (1980) 245\\
A.V. Efremov, A.V. Radyushkin, Teor. i Mathem. Fiz. {\bf 42} (1980) 147
\bibitem{LB}
G.P. Lepage, S.J. Brodsky, Phys. Rev. {\bf D22} (1980) 2157
\bibitem{cz-r}
V.L. Chernyak, A.R. Zhitnitsky, Phys. Rep. {\bf 112} (1984) 173-318
\bibitem{BL}
S.J. Brodsky, G.P. Lepage, Phys. Rev. {\bf D24} (1981) 1808
\bibitem{Maurice}
M. Benayoun, V.L. Chernyak, Nucl. Phys. {\bf B329} (1990) 285
\bibitem{cz2}
V.L. Chernyak, A.R. Zhitnitsky, Nucl. Phys. {\bf B201} (1982) 492 ,\\
Erratum: {\it ibid} {\bf B214} (1983) 547
\bibitem{Nakaz}
H. Nakazawa et al., Belle Collaboration, Measurement of the $\gamma\gamma\rightarrow \pi^{+}\pi^{-}$
and $\gamma\gamma\rightarrow K^{+}K^{-}$ processes at energies $2.4-4.1\,GeV$,
Phys. Lett. {\bf B615} (2005) 39, hep-ex/0412058
\bibitem{czz}
V.L. Chernyak, A.R. Zhitnitsky, I.R. Zhitnitsky, Nucl. Phys. {\bf B204} (1982) 477
\bibitem{Chen}
W.T. Chen et al., Belle Collaboration, A study of $\gamma\gamma\rightarrow K_SK_S$ production at
energies of $2.4-4.0\, GeV$ at Belle, Phys. Lett. {\bf B651} (2007) 15, hep-ex/0609042
\bibitem{Ue-pi}
S. Uehara, Y. Watanabe et al., Belle Collaboration, High statistic measurement of neutral-pion pair
production in two-photon collisions, Phys. Rev.,  {\bf D79} (2009) 052009, arXiv: 0903.3697 [hep-ex]
\bibitem{Ue-eta}
S. Uehara, Y. Watanabe et al., Belle Collaboration, High statistic study of $\eta\pi^{o}$ production
in two-photon collisions, Phys. Rev., {\bf D80} (2009) 032001\,, arXiv: 0906.1464 [hep-ex]
\bibitem{Ue-ee}
S. Uehara, Y. Watanabe\, et\, al., Belle Collaboration, \,Phys. Rev., {\bf D82} (2010) 114031
\bibitem{GI}
I.F. Ginzburg, D.Yu. Ivanov, Nucl. Phys. {\bf B388} (1992) 376
\bibitem{DKV}
M. Diehl, P. Kroll, C. Vogt, Phys. Lett. {\bf B532} (2002) 99,  hep-ph/0112274
\bibitem{Ch1}
V.L. Chernyak, Phys. Lett. {\bf B640} (2006) 246,  hep-ph/0605072
\bibitem{Ch2}
V.L. Chernyak, Nucl. Phys. {\bf} (Proc. Suppl.) {\bf 162} (2006) 161,  hep-ph/0605327
\bibitem{Braun}
I.I. Balitsky, V.M. Braun, A.V. Kolesnichenko, Nucl. Phys. {\bf B312} (1989) 509
\bibitem{cz3}
V.L. Chernyak, I.R. Zhitnitsky, Nucl. Phys. {\bf B345} (1990) 137
\bibitem{Ch3}
V.L.\, Chernyak, arXiv\,:\,\,0912.0623\,\,[hep-ph]
\bibitem{DN}
G. Duplancic, B. Nizic, Phys. Rev. Lett., 97 (2006) 142003
\bibitem{Duplan}
G. Duplancic, Private communication
\bibitem{DK}
M.\, Diehl, P.\, Kroll, Phys. Lett. {\bf B683} (2010) 165
\bibitem{cz4}
V.L. Chernyak, I.R. Zhitnitsky, Nucl. Phys. {\bf B222} (1983) 382
\bibitem{GPS}
I.F. Ginzburg, S.L. Panfil, V.G. Serbo, Nucl. Phys. {\bf B284} (1987) 685
\bibitem{LSY}
Z.Q. Liu, C.P. Shen, C.Z. Yuan et al., Belle\, Collaboration,\\ Phys. Rev. Lett.,\,
{\bf 108}, 232001 (2012),
\, arXiv: 1202.5632 [hep-ex]
\bibitem{LSY1}
Z.Q. Liu, C.P. Shen, C.Z. Yuan,\, Private communication

\end{thebibliography}
\end{document}